\documentclass[11pt,a4paper]{article}
\pdfoutput=1
\usepackage{jcappub}
\usepackage{amsmath,amssymb}
\usepackage{epsfig,graphicx}
\begin{document}

\title{Possible evolution of a bouncing universe in cosmological models with non-minimally coupled scalar fields}

\author[a]{Ekaterina~O.~Pozdeeva,}
\affiliation[a]{Skobeltsyn Institute of Nuclear Physics,  Lomonosov
Moscow State University, Leninskie Gory~1, 119991, Moscow, Russia}
 \emailAdd{pozdeeva@www-hep.sinp.msu.ru}
 \author[b]{Maria~A.~Skugoreva,}
\affiliation[b]{Kazan Federal University, Kremlevskaya str.~18, Kazan, 420008, Russia}
 \emailAdd{masha-sk@mail.ru}
  \author[b,c]{Alexey~V.~Toporensky,}
\affiliation[c]{Sternberg Astronomical Institute, Lomonosov Moscow State University,\\
\small Universitetsky pr.~13, Moscow, 119991, Russia}
 \emailAdd{atopor@rambler.ru}
\author[a]{Sergey~Yu.~Vernov}
\emailAdd{svernov@theory.sinp.msu.ru}

\date{ \ }

\abstract{
  We explore dynamics of cosmological models with bounce solutions evolving on a spatially flat Friedmann--Lema\^{i}tre--Robertson--Walker background. We consider cosmological models that contain the Hilbert--Einstein curvature term, the induced gravity term with a negative coupled constant, and   even polynomial potentials of the scalar field. Bounce solutions with non-monotonic Hubble parameters have been obtained and analyzed.
  The case when the scalar field has the conformal coupling and the Higgs-like potential with an opposite sign is studied in detail.
 In this model the evolution of the Hubble parameter of the bounce solution essentially depends on the sign of the cosmological constant.}


\arxivnumber{1608.08214}


\maketitle

\section{Introduction}

The global evolution of the observable Universe can be separated on four epochs: inflation, radiation domination era, matter domination era and dark energy domination era. All these epochs can be described by General Relativity models with minimally coupled scalar fields, because the assumption that the Hubble parameter is a monotonically decreasing function does not contradict to the observation data. In such type of models the initial period of the Universe evolution with energies above the Planck scale should be described by quantum gravity because the classical evolution includes the initial singularity. Important question of theoretical cosmology is whether the entire Universe evolution can remain classical and has no singularity.

It is possible to avoid this singularity considering modified gravity. Bouncing universe smoothly transits from a period of contraction to a period of expansion and its evolution remains classical.
 Models of bouncing universes attract a lot of attention~\cite{Novello:2008ra,Battefeld:2014uga,Brandenberger:2016vhg}. We can mention, in particular, bouncing models with Galileon fields~\cite{Qiu:2011cy,GBounce,Osipov:2013ssa,Koehn:2013upa,Libanov:2016kfc,Kobayashi:2016xpl,Ijjas:2016tpn}, $f(R)$ gravity models~\cite{Bamba:2013fha,Nojiri:2016ygo}, non-local gravity models~\cite{Biswas:2005qr,Biswas:2012bp,Koshelev:2012qn,Koshelev:2013lfm,Calcagni:2013vra} and models with non-minimal coupling~\cite{Tretyakova2011,Boisseau:2015hqa,KPTVV2015,Boisseau:2016pfh}.

The goal of this paper is to explore new mathematical features of cosmological
models with non-minimally coupled scalar fields that admit bounce solutions. We consider models with the spatially flat Friedmann--Lema\^{i}tre--Robertson--Walker (FLRW) metric\footnote{Note that the bouncing models with the open or close FLRW metric are also known~\cite{Star78,MO79,Star81,Ferreira:2015iaa}.}.  At the bounce point the period of universe contraction changes to a period of universe expansion. Thereby, a bounce point is characterized by two conditions: at this point the Hubble parameter $H$ is equal to zero and its cosmic time derivative is positive.

Note that the General Relativity models with  minimally coupled standard scalar fields have no bounce solution, because in these models the Hubble parameter  does not increase. The simplest way to get a bounce solution is to consider a model with a phantom scalar field. Point out that similar models are used to describe dark energy with the state parameter less then minus one~\cite{Dabrowski:2003jm,Elizalde:2004mq,Phantom_models}. To get non-monotonic behaviour of the Hubble parameter quintom models have been developed. These  models can describe both bounce~\cite{QUINTOMBOUNCE}, and dark energy~\cite{QUINTOM,AKV2}. In models with  phantom fields the Null Energy Condition is violated and instability problems arise~\cite{Rubakov:2014jja}. At the same time it is easy to get increasing  or non-monotonic behavior of the Hubble parameter in models with a standard scalar field non-minimally coupled to gravity~\cite{Gannouji:2006jm}, for example, in induced gravity models~\cite{KTVV2013}. Such scalar-tensor models
have no ghost~\cite{Gannouji:2006jm,EspositoFarese:2000ij}.

To obtain bounce solutions in  cosmological models with standard scalar fields one can use non-minimal coupling between the scalar fields and gravity
and consider the following action:
\begin{equation}
\label{S1}
S =\int d^4x\sqrt{-g}\left(U(\varphi)R-\frac{1}{2}g^{\mu\nu}\partial_{\mu}\varphi\partial_{\nu}\varphi- V(\varphi)\right ),
\end{equation}
where $g=\text{det}(g_{\mu\nu})$ is the determinant consisting of metric tensor components $g_{\mu\nu}$, $R$ is the Ricci scalar,
$U(\varphi)$ and $V(\varphi)$ are differentiable functions of a scalar field $\varphi$.

From the condition that at the bounce point the Hubble parameter $H=0$, it follows that  the potential should be negative: $V(\varphi_b)<0$.
The characteristic property of models with potentials that are not positive definite is the existence of the unreachable domain on the phase plane, which
corresponds to non-real values of the Hubble parameter~\cite{Felder:2002jk,Giambo:2014jfa,ABG,ABGV,Giambo:2015tja}.

In this paper we consider models with bounce solutions that generalize the integrable model proposed in~\cite{Boisseau:2015hqa} with the Ricci scalar $R$ as an integral of motion. This model has the potential in the form $V(\varphi)=C_0+C_4\varphi^4$, with $C_0>0$ and $C_4<0$, and the standard quadratic coupling $U(\varphi)=U_0-\xi \varphi^2/2$ with  $\xi=1/6$ and a positive constant $U_0$. As known, in the spatially flat FLRW metric the Ricci scalar $R$ is a function of the Hubble parameter $H$ and its first derivative. So, the condition that  $R$ is a constant defines the Hubble parameter as a solution of the first order differential equation. Thereby, we get only monotonic behavior of the Hubble parameter (formulae are presented in Section 3). In particular, this integrable model contains bouncing
cosmological solutions with smooth future behavior tending to a de Sitter solution (while past behavior shows emerging of a Universe
at a point with zero effective Newtonian constant). It has been shown in~\cite{KPTVV2015} that the integrable cosmological model~\cite{Boisseau:2015hqa} with a constant $R$ belongs to one-parameter set of integrable models with one and the same function $U$ and different potentials $V$. However, the Ricci scalar $R$ is not an integral to motion for these integrable models. In this paper we show that the model, proposed in~\cite{Boisseau:2015hqa}, is unique in the sense that  the property of
having $R$ to be a constant disappears, when one changes coupling function or scalar field potential (or both).

To construct more realistic models with both a bounce, and a non-monotonic behavior of the Hubble parameter one can consider non-integrable models that are close to the above-mention integrable model with a constant $R$. In~\cite{Boisseau:2016pfh} the authors got such types of the Hubble parameter considering non-integrable models with the same potential and other values of $\xi$. In the present paper we mostly vary the potential in searching for conditions of
bouncing behavior similar to studied in~\cite{Boisseau:2015hqa} to exist in these more general models. In Section~4 we consider the model with $\xi=1/6$ and the Higgs-like potential multiplying by a negative constant. In Section~5 we consider models with different positive values of $\xi$, a cosmological constant and different even degrees of monomial potentials. Using numerical calculations we demonstrate the possibility of different evolutions in the cases of potentials of the second, fourth and sixth degrees.
We analyze the reasons of such behavior and find the corresponding conditions on the model parameters.

The paper is organized as follows. In Section~2 we remind a reader the general bounce conditions in non-minimally coupled scalar field theory as well as the conception of the effective potential. In Section~3 we show that the choice of potential in the paper~\cite{Boisseau:2015hqa} leading to the property that the curvature $R$ is the integral of motion is the unique one. On the contrary, in Sections~4 and 5 we show that bounce behavior admits much larger class of potentials. In Section~4 bounce behavior is studied for potentials with quadratic term in addition to quartic and constant terms and $\xi=1/6$.  The models with $\xi\neq 1/6$ are considered in Section~5. In Section~6 we show that the obtained bouncing solutions do not suffer from the gradient or ghost instability. Section~7 contains a summary of the result obtained.

\section{Bounce solutions in non-minimally coupled models }

In the spatially flat FLRW metric with the interval:
\begin{equation*}
ds^2={}-dt^2+a^2(t)\left(dx_1^2+dx_2^2+dx_3^2\,\right),
\end{equation*}
we obtain from action (\ref{S1}) the following equations:
\begin{equation}
\label{Frequoc00}
6UH^2+6\dot UH -\frac12\dot{\varphi}^2-V=0,
\end{equation}
\begin{equation}
\label{Frequocii}
 2U\left[2\dot{H}+3H^2\right]
+2U'\left[\ddot{\varphi}+2H\dot{\varphi}\right] =V-\left[2U''+\frac12\right]
\dot{\varphi}^2,
\end{equation}
\begin{equation}
\ddot{\varphi}+3H\dot{\varphi} -6U'\left[\dot{H}+2H^2\right]+V' = 0,
\label{KGoc}
\end{equation}
where a ``dot'' means a derivative with respect to the cosmic time $t$, and a ``prime'' means a derivative with respect to the scalar field $\varphi$.
The function $a(t)$ is the scale factor, its logarithmic derivative $H= \dot{a}/a$ is the Hubble parameter. We note that an effective gravitation constant in the model considered is
\begin{equation}
\label{Geff}
G_{eff}=\frac{1}{16\pi U}.
\end{equation}
The dynamics of the FLRW Universe can be prolonged smoothly into the region of $G_{eff}<0$ (see, for example~\cite{Skugoreva:2012,KPTVV2016}), however,  any anisotropic or inhomogeneous corrections are expected to diverge while $G_{eff}$ tends to infinity~\cite{Star81,Caputa:2013mfa}. In this paper we analyze such bounce solutions that  $U(\varphi(t))>0$ for any $t\geqslant t_b$ and conditions of their existence.

If a solution of Eqs.~(\ref{Frequoc00})--(\ref{KGoc}) has such a point $t_b$ that $H(t_b)=0$ and $\dot{H}(t_b)>0$, then it is a bounce solution. Let us find conditions that are necessary for the existence of a bounce in models with action~(\ref{S1}).
Using Eq.~(\ref{Frequoc00}), we get that from $H(t_b)=0$ it follows  $V(\varphi(t_b))\leqslant 0$.
Subtracting equation~(\ref{Frequoc00}) from equation~(\ref{Frequocii}), we obtain
\begin{equation}
\label{Frequ21}
   4U\dot H={}-{\dot\varphi}^2-2\ddot U+2H\dot U.
\end{equation}

From Eq.~(\ref{Frequ21}) it follows that a bounce solution does not exist if $U$ is a positive constant.
Equation~(\ref{Frequ21})  at the bounce point gives
\begin{equation}
\label{bounceCond2}
2\left(U+3{U'}^2\right)\dot{H}(t_b)=U'V'+\left[2U''+1\right]V,
\end{equation}
where functions $U$ and $V$ and their derivatives are taken at the point $\varphi(t_b)$.
The condition $\dot{H}(t_b)>0$ gives the restriction on functions $U$ and $V$ at the bounce point.

To analyze the dynamic it is suitable to introduce a new variable and present equations (\ref{Frequoc00}) and (\ref{Frequ21})
in the form similar to the Friedmann equations in the Einstein frame. Following~\cite{Skugoreva:2014gka}, we introduce the effective potential
\begin{equation}
\label{Veff}
V_{eff}(\varphi)=\frac{V(\varphi)}{4K^2 U(\varphi) ^2}.
\end{equation}
If the constant $K=8\pi G$, then $V_{eff}$ coincides with the potential of the corresponding model in the Einstein frame. Note that we do not transform the metric and consider the model in the Jordan frame only. For our purposes any positive value of $K$ is suitable (in~\cite{Skugoreva:2014gka} it was chosen $K=6$). Also in~\cite{Skugoreva:2014gka}  functions
\begin{equation}
\label{PA}
P\equiv \frac{H}{\sqrt{U}}+\frac{U'\dot\varphi}{2U\sqrt{U}},\qquad A\equiv\frac{U+3{U'}^2}{4U^3}
\end{equation}
have been introduced. If $U(\varphi)>0$, then $A(\varphi)>0$ as well.

In terms of these
 functions Eqs.~(\ref{Frequoc00}) and (\ref{Frequ21}) take the following form:
\begin{equation}
\label{equP}
3P^2=A{\dot\varphi}^2+2K^2V_{eff},
\end{equation}
\begin{equation}
\label{Fr21Qm}
\dot P={}-A\sqrt{U}\,{\dot\varphi}^2.
\end{equation}

As known the Hubble parameter is a monotonically decreasing function in  models with a standard scalar field minimally coupled to gravity ($U(\varphi)$ is a constant). For a model with an arbitrary positive $U(\varphi)$ the function $P(\varphi)$ has the same property.

Equation~(\ref{Frequoc00}) is a quadratic equation in $H(t)$ that has the following solutions:
\begin{equation}\label{hpm}
    H_{\pm}={}-\frac{\dot{U}}{2U}\pm\frac{1}{6U}\sqrt{9{U'}^2\dot{\varphi}^2+3U\dot{\varphi}^2+6UV}.
\end{equation}

The value of the function $P(\varphi)$ that correspond to $H_{\pm}$ is
\begin{equation}\label{Ppm}
    P=\pm\frac{1}{6U}\sqrt{3\left[\frac{3{U'}^2}{U}\dot{\varphi}^2+\dot\varphi^2+2V\right]}
    =\pm\sqrt{\frac{A}{3}{\dot\varphi}^2+\frac{2}{3}K^2V_{eff}} \, ,
\end{equation}
So, a positive $P$ corresponds to $H_+$ and a negative $P$ corresponds to $H_-$.

If the potential $V$ is negative at some values of $\varphi$, whereas the function $U$ is positive at these points, then  one should restrict the domain of absolute values of~$\dot\varphi$ from below to get a real Hubble parameter. In other words, there exists the unreachable domain on the phase plane~\cite{ABGV}. The boundary of this domain is defined by the condition $P=0$. If $V_{eff}>0$, then the sign of the function $P$ cannot be changed. If in some domain $V_{eff}(\varphi)$ is negative, then the sign of $P$ can be changed from plus to minus, but not vice verse. So, the sign of $P$ cannot be changed twice.

Equations~(\ref{Frequoc00})--(\ref{KGoc}) can be transformed into the following system of the first order equations which is useful for  numerical calculations and analysis of stability:
\begin{equation}
\left\{
\begin{split}
\dot\varphi&=\psi,\\
\dot\psi&={}-3H\psi-\frac{\left[(6 U''+1)\psi^2-4V\right]U'+2UV'}{2\left(3 {U'}^2+ U\right)},\\
\dot H&={}-\frac{2U''+1}{4\left(3{U'}^2+U\right)}\psi^2+\frac{2U'}{{3{U'}^2+U}}H\psi
-\frac{6{U'}^2}{3{U'}^2+U}H^2+\frac{U'V'}{2\left(3{U'}^2+U\right)}\,.
\end{split}
\right.
\label{FOSEQU}
\end{equation}
If Eq.~(\ref{Frequoc00}) is satisfied in the initial moment of time, then from system (\ref{FOSEQU}) it follows
that Eq.~(\ref{Frequoc00}) is satisfied at any moment of time. So, one can use Eq.~(\ref{Frequoc00}) to fix initial conditions of system~(\ref{FOSEQU}).
From Eqs.~(\ref{equP}) and (\ref{Fr21Qm}) it is easy to get the following system~\cite{Skugoreva:2014gka}:
\begin{equation}
\label{system2}
\dot\varphi=\psi\,,\qquad
\dot\psi={}-3P\sqrt{U}\psi-\frac{A'}{2A}\psi^2-\frac{K^2V'_{eff}}{A}.
\end{equation}
System of equations (\ref{Fr21Qm}) and (\ref{system2}) is equivalent to system (\ref{FOSEQU}) if $U(\varphi)>0$ for any $\varphi$. On the other hand, at $U=0$ the functions $P$, $A$ and $V_{eff}$ are singular, whereas system (\ref{system2}) has no singularity at this point.

De Sitter solutions correspond to $\psi=0$, and hence,
$V'_{eff}(\varphi_{dS})=0$. The corresponding Hubble parameter is
\begin{equation}
\label{Hf}
H_{dS}=P_{dS}\sqrt{U(\varphi_{dS})}=\pm \sqrt{\frac{2}{3}K^2U(\varphi_{dS})V_{eff}(\varphi_{dS})}
=\pm\sqrt{\frac{V(\varphi_{dS})}{6U(\varphi_{dS})}}\,.
\end{equation}

We analyze the stability of de Sitter solutions with $H_{dS}>0$ and $U(\varphi_{dS})>0$ only. From (\ref{Hf}) it follows that  $V(\varphi_{dS})>0$. In this case the Hubble parameter in the neighborhood of $H_{dS}$ is uniquely defined by Eq.~(\ref{Frequoc00}), so it is enough to consider system (\ref{system2}) to analyze the stability of de Sitter solution.
Substituting the de Sitter solution with the first order perturbations:
\begin{equation}
\varphi(t)=\varphi_{dS}+\varphi_1(t),\qquad \psi(t)=\psi_1(t),
\end{equation}
 into (\ref{system2}), we get the following linear system on $\varphi_1(t)$ and $\psi_1(t)$:
\begin{equation}
\label{linsystem}
\begin{split}
\dot\varphi_1&=\psi_1,\\
\dot\psi_1&={}-K^2\frac{V''_{eff}(\varphi_{dS})}{A(\varphi_{dS})}\varphi_1-3H_{dS}\psi_1.
\end{split}
\end{equation}

We find eigenvalues for system (\ref{linsystem}):
\begin{equation}
\label{eigen}
\begin{array}{l}
\begin{vmatrix}
-\lambda & 1\\
-\frac{K^2V''_{eff}(\varphi_{dS})}{A(\varphi_{dS})} & -3H_{dS}-\lambda
\end{vmatrix}
=\lambda^2 +3H_{dS}\lambda+\frac{K^2V''_{eff}(\varphi_{dS})}{A(\varphi_{dS})}=0 \Rightarrow\\
\\\Rightarrow\displaystyle \lambda_\pm=-\frac{3}{2}H_{dS}\pm\frac{1}{2}\sqrt{9{H_{dS}}^2-\frac{4 K^2 V''_{eff}(\varphi_{dS})}{A(\varphi_{dS})}}.
\end{array}
\end{equation}

The real part of $\lambda_-<0$ always, the condition that the real part of $\lambda_+$ is negative is equivalent to $V''_{eff}(\varphi_{dS})>0$.
    Therefore, for arbitrary differentiable functions  $V$ and $U>0$,  the model has a stable de Sitter solution with $H_{dS}>0$  only if the potential $V_{eff}$ has a minimum~\cite{Skugoreva:2014gka} and $V_{eff}(\varphi_{dS})>0$.

Using the obtained values of $\lambda_\pm$, it is easy to find~\cite{Tabor} that the de Sitter point is a stable node (the scalar field decreases monotonically)
at
\begin{equation}
\label{node}
\frac{3\left(U+3{U'}^2\right)}{8U^2}\geqslant \frac{V''_{eff}}{V_{eff}},
\end{equation}
and a stable focus (the scalar field oscillations exist)
at
\begin{equation}
\label{focus}
\frac{3\left(U+3{U'}^2\right)}{8 U^2}<\frac{V''_{eff}}{V_{eff}}.
\end{equation}

\section{The choice of the function $U$}

In the paper~\cite{Boisseau:2015hqa}  the cosmological model with a constant Ricci scalar $R$ has been considered.
In this section we demonstrate how the form of function $U$ and the corresponding potential can be found from the requirement that the Ricci scalar is an integral of motion.

In the spatially flat FLRW metric $R=6(\dot H+2H^2)$. From Eqs.~(\ref{Frequoc00})--(\ref{KGoc}) we get
\begin{equation}
\label{equtrace}
2UR=-{\dot {\varphi}}^{2} +4
V-6 (3 H \dot{\varphi}  U'+ {\dot
{\varphi}}^{2}U''+\ddot {\varphi} U'),
\end{equation}
\begin{equation}
\label{KGequ}
\ddot\varphi +3 H\dot\varphi-U'R+V'=0.
\end{equation}
 Combining Eqs.~(\ref{equtrace}) and (\ref{KGequ}), we obtain

\begin{equation}
\label{equR}
2R\left( U+3{U'}^2 \right) +\left(6 U''+1\right) {\dot {\varphi}}^{2}=4V
+6 V'U'\,.
\end{equation}

From the structure of Eq.~(\ref{equR}) it is easy to see that the simplest way to get a constant $R$ is to choose such $U(\varphi)$ that
\begin{equation}
\label{eqU}
U+3 U'^2=U_0,\qquad 6 U''+1=0, \qquad U_0R=2V+3 V'U',
\end{equation}
where $U_0$ is a constant.

The first two equations of system~(\ref{eqU})  have the following solution
\begin{equation}
\label{Uc}
    U_c(\varphi)=U_0-\frac{1}{12}(\varphi-\varphi_0)^2,
\end{equation}
where $\varphi_0$ is an integration constant. Without loss of generality we put $\varphi_0=0$.
For such a choice of $U(\varphi)$ we get Eq.~(\ref{equR}) as follows:
\begin{equation}
\label{equR2}
2U_0R=4V(\varphi)-\varphi V'(\varphi).
\end{equation}
Considering Eq.~(\ref{equR2}) as a differential equation for $V(\varphi)$, for a constant $R$ we get the following solution:
\begin{equation}
\label{V4}
V_{int}=\frac{\Lambda}{K}+C_4\varphi^4,
\end{equation}
where $C_4$ is an integration constant, for other constants we choose
\begin{equation}
\Lambda=\frac{R}{4}, \qquad K=\frac{1}{2U_0}.
\end{equation}

Thus, requiring that the Ricci scalar is a constant, one can define both functions $U(\varphi)$ and $V(\varphi)$. To get a positive $G_{eff}$ for some values of $\varphi$ we choose $U_0>0$.

 Substituting $U_c$ and $V_{int}$ into Eq.~(\ref{bounceCond2}), we get that the condition $\dot{H}_b>0$ is equivalent to $\Lambda>0$, hence, from $V(\varphi_b)<0$ it follows $C_4<0$. This integrable cosmological model  has been considered in~\cite{Boisseau:2015hqa}, where the behavior of bounce solutions has been studied in detail.

Equation $R=4\Lambda$ is a differential equation for the Hubble parameter:
\begin{equation}
\label{equHR}
3\left(\dot H+2H^2\right)=2\Lambda,\qquad \Leftrightarrow \qquad
6\left(\ddot a a+\dot a^2\right)=4\Lambda a^2.
\end{equation}
Multiplying this equation by $a\dot a$, we get
\begin{equation}
6\ddot a\dot a a^2+6\dot a^3 a - 4\Lambda a^3\dot a=\frac{d}{dt}\left[3\dot{a}^2 a^2-\Lambda a^4\right]= 0.
\end{equation}
Therefore, there exists the following integral of motion:
\begin{equation}
3\dot{a}^2 a^2-\Lambda a^4=C.
\end{equation}

Equation~(\ref{equHR}) with a positive $\Lambda$ has two possible real solutions in dependence of the initial conditions:
\begin{equation}
\label{Hmonot}
H_1(t) = \sqrt{\frac{\Lambda}{3}}\tanh\left(\frac{2}{3}\sqrt{3\Lambda}(t-t_0)\right),\qquad H_2(t) = \sqrt{\frac{\Lambda}{3}}\coth\left(\frac{2}{3}\sqrt{3\Lambda}(t-t_0)\right),
\end{equation}
where $t_0$ is an integration constant. Note that the behavior of the Hubble parameter does not depend on the specific dynamics of the scalar field $\varphi$, because two-parametric set of functions $\varphi(t)$  corresponds to one-parametric set of $H(t)$.

\section{Models with fourth-degree even polynomial potentials}

\subsection{Equations of motion and restrictions on parameters}
The monotonically increasing Hubble parameter $H_1(t)$ is not suitable for construction of a realistic cosmological scenario, whereas
$H_2(t)$ is not a bounce solution. In~\cite{Boisseau:2016pfh} the authors slightly modify the function $U$ to get a bounce solution with non-monotonic behaviour of the Hubble parameters. In this section we modify the potential instead of the function $U$ and analyze the obtained bounce solutions. In other words, we choose the function $U=U_c$, given by (\ref{Uc}), with $U_0=1/(2K)$. As a minimal generalization of the potential $V_{int}$ we consider the following potential
\begin{equation}\label{V4m}
    V_c=C_4\varphi^4+C_2\varphi^2+C_0,
\end{equation}
where $C_i$ are constants. We consider the case $C_4<0$ only, because the integrable model with the potential $V_{int}$ has a bounce solution and a stable de Sitter solution only if $\Lambda>0$ and $C_4<0$.

We plan to consider the evolution of bounce solutions fixing the initial conditions at the bounce point. Without loss of generality
we can consider an initial value $\varphi_i>0$ only. The initial value of $\psi$ is defined by the condition $H_i=0$, hence, $\psi_i^2=-2V_c(\varphi_i)$.

Our first goal is to find conditions on the coefficients of the potential $V_c$ that correspond to the existence of bounce solutions.
For  $U=U_c$, a bounce solution exists only if there exists such a point $\varphi_b$ that the following conditions are satisfied~\cite{KPTVV2015}:
\begin{equation}
V(\varphi_b)<0,\qquad 4V(\varphi_b)-\varphi_b V^\prime(\varphi_b)>0.\label{bounceCondition}
\end{equation}
So, $V^\prime(\varphi_b)<0$, and we obtain the following conditions on parameters $C_i$:
\begin{equation}
\label{Bouncecond}
  C_4\varphi_b^4+C_2\varphi_b^2+C_0<0, \qquad  C_2\varphi_b^2+2C_0>0, \qquad
C_2+2C_4\varphi_b^2<0.
\end{equation}
Thus, at least one of the constants $C_2$ or $C_0$ should be positive.
A bounce solution has the physical sense only if $G_{eff}(\varphi_b)>0$. The condition $U_c(\varphi_b)>0$ means  $\varphi_b^2<6/K$.

The effective potential (\ref{Veff}) is
\begin{equation}\label{Veff4}
V_{eff}=\frac{36(C_4\varphi^4+C_2\varphi^2+C_0)}{(K\varphi^2-6)^2}.
\end{equation}
The even potential $V_{eff}$ has an extremum at $\varphi=0$ and at points
\begin{equation}
\label{phiextr}
    \varphi_m=\pm\sqrt{\frac{-2(3 C_2+K C_0)}{12 C_4+K C_2}}.
\end{equation}
We consider such values of parameter that points $\varphi_m$ are real and $0<\varphi_m^2<6/K$.
Note that for integrable model ($C_4<0$,  $C_2=0$, $C_0>0$)  $ \varphi_m$ are non-zero real numbers  and the condition $\varphi_m^2<6/K$  is equivalent to  $-C_0/C_4<(6/K)^2$.

From~(\ref{Bouncecond}),  $C_4<0$ and $\varphi_b^2<6/K$ we get
\begin{equation}
0>C_2+2\varphi_b^2C_4> C_2+\frac{12}{K}C_4.
\end{equation}
So, the model with a bounce solution has real $\varphi_m$ only at
\begin{equation}\label{condvarphim}
    3 C_2+K C_0>0 \qquad\mbox{and}\qquad KC_2+12C_4<0.
\end{equation}
Using these conditions, we get
\begin{equation*}
V''_{eff}(0)=2\left(\frac{1}{3}C_0K+C_2\right)>0,\qquad V''_{eff}(\varphi_m)={}-\frac{36(C_2K+12C_4)^3(C_0K+3C_2)}{(C_0K^2+6C_2K+36C_4)^3}<0.
\end{equation*}
so, the potential $V_{eff}$ has a minimum at $\varphi=0$ and maxima at $\varphi=\varphi_m$.

\begin{figure}[!h]
\centering
\includegraphics[width=45mm]{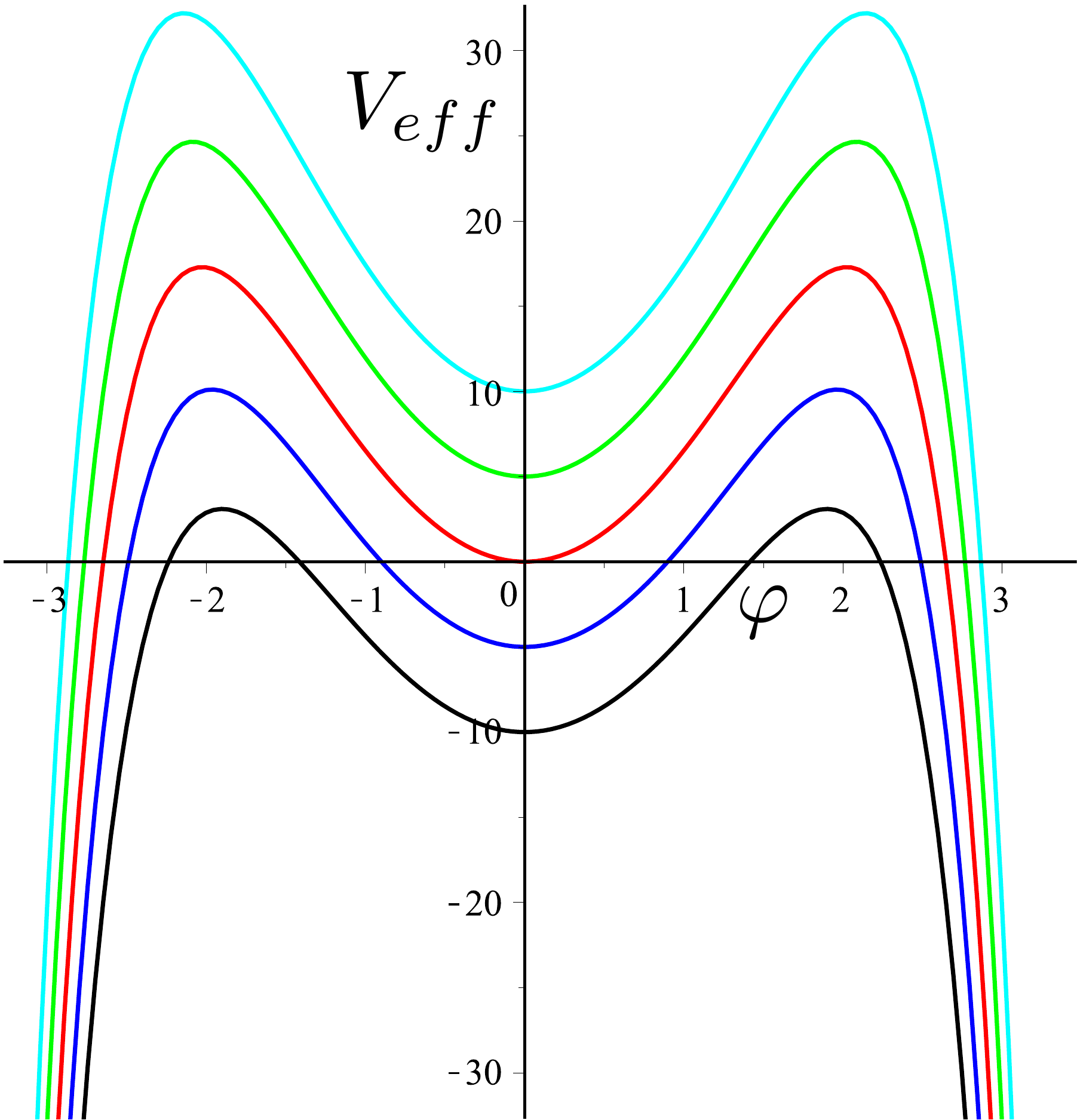}   \qquad
\includegraphics[width=45mm]{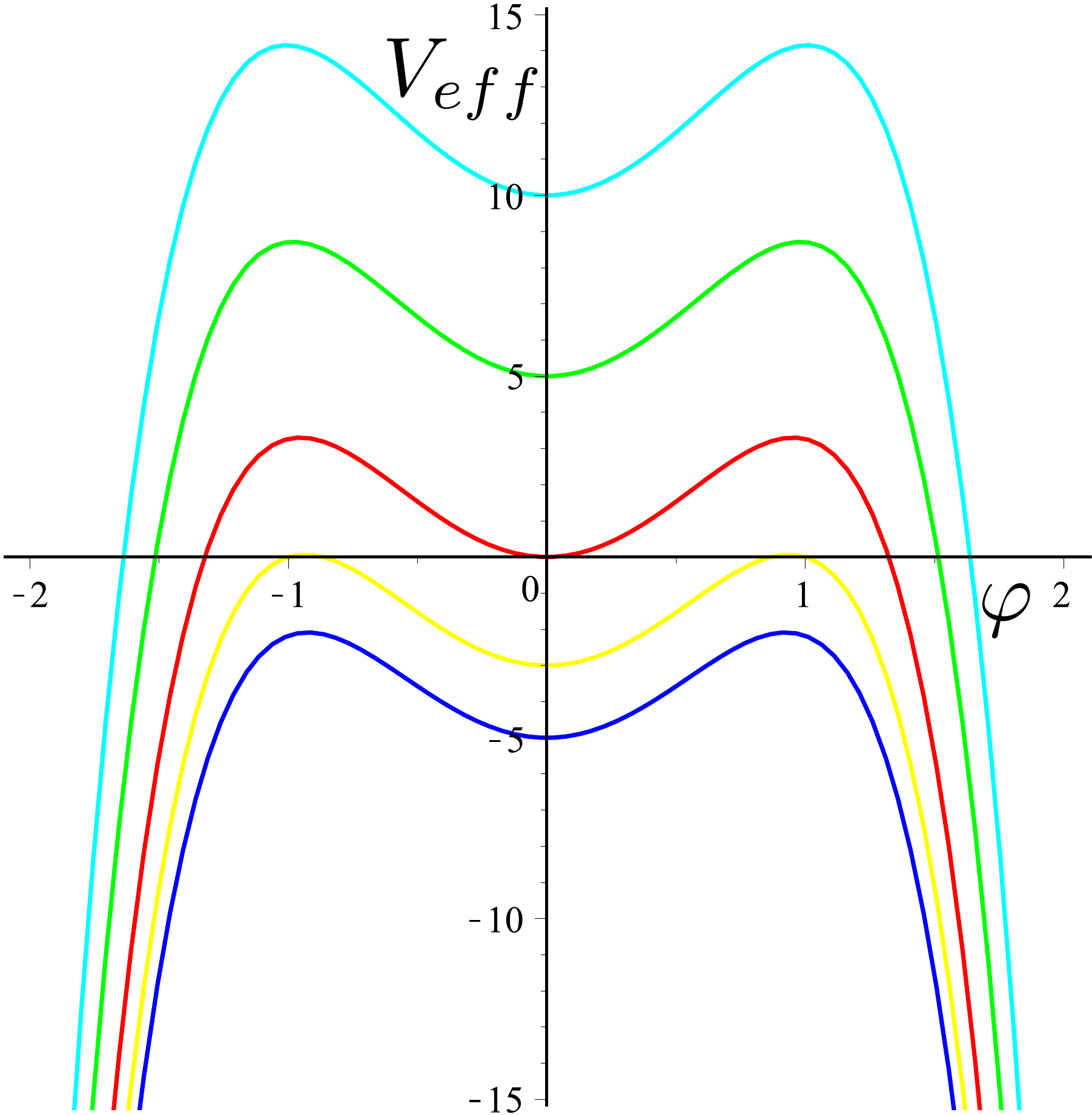}  \qquad
\includegraphics[width=45mm]{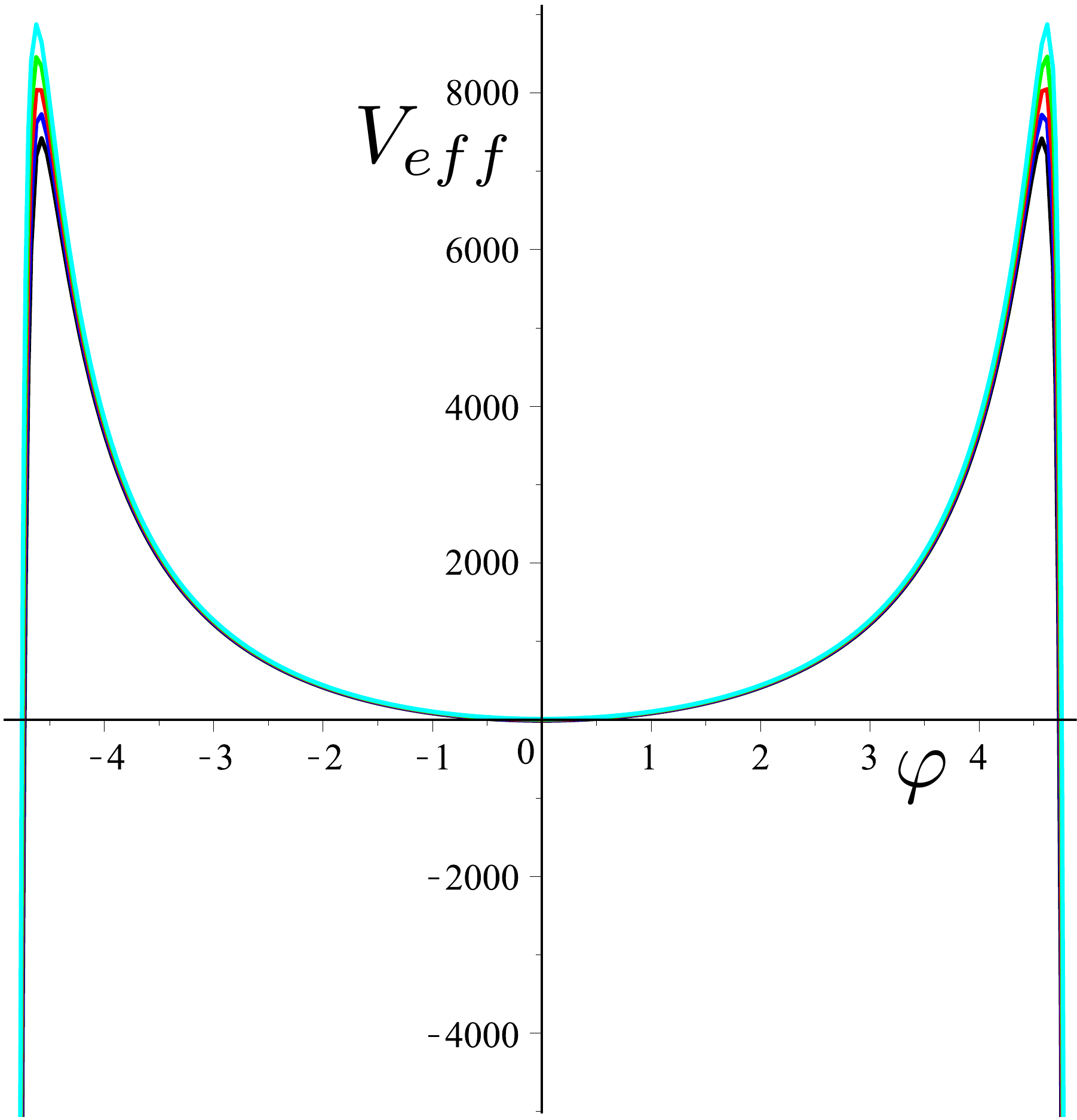}
\caption{The effective potential $V_{eff}$ at different values of parameters. In all pictures we choose $K=1/4$. The values of parameters are
 $C_4=-1$,  $C_2= 7$ (left picture),  $C_4=-4$,  $C_2= 7$ (middle picture),  $C_4=-4$,  $C_2= 90$ (right picture). The parameter $C_0=-10$ (black curve),  $-5$ (red curve),   $0$ (blue curve), $5$ (green curve), and  $10$ (cyan curve). The yellow curve in the middle picture corresponds to $C_0=-3$.}
\label{VeffC0}
\end{figure}

A few examples of the effective potential $V_{eff}$ with two maxima at nonzero points $\varphi_m$ and a minimum at $\varphi=0$ are presented in Fig.~\ref{VeffC0}.

Stable de Sitter solution at $\varphi_{dS}=0$ corresponds to $H_{dS}^2=C_0 K/{3}$. Therefore, such a solution exists for $C_0\geqslant0$ only. It means that the stable de Sitter solution exists only if $V_{eff}(0)\geqslant0$. In this case $V_{eff}(\varphi_m)>0$.

The potential   $V_{eff}$ has the following zeros:
\begin{equation*}
  \varphi_1^2 = \frac12\left(\sqrt{\left(\frac{C_2}{C_4}\right)^2-4\frac{C_0}{C_4}}-\frac{C_2}{C_4}\right), \qquad
  \varphi_2^2={}-\frac12\left(\sqrt{\left(\frac{C_2}{C_4}\right)^2-4\frac{C_0}{C_4}}+\frac{C_2}{C_4}\right).
\end{equation*}

At $C_4<0$, $C_0>0$ one gets $\sqrt{\left(\frac{C_2}{C_4}\right)^2-4\frac{C_0}{C_4}}>\frac{|C_2|}{|C_4|}$. Therefore, $\varphi_2^2<0$
and there exist only two real roots:
\begin{equation*}
  \varphi_1^{-} = -\sqrt{\frac12\left(\sqrt{\left(\frac{C_2}{C_4}\right)^2-4\frac{C_0}{C_4}}-\frac{C_2}{C_4}\right)}\,, \qquad
  \varphi_1^{+} = \sqrt{\frac12\left(\sqrt{\left(\frac{C_2}{C_4}\right)^2-4\frac{C_0}{C_4}}-\frac{C_2}{C_4}\right)}.
\end{equation*}
The bounce point $\varphi_b>0$ belongs to the interval $\varphi\in\left(\varphi_1^{+},\sqrt{6/K}\right)$.
We get the condition $\varphi_1^{+}<\sqrt{6/K}$.

\subsection{Analysis of numeric solutions}

For $U=U_c$ and an arbitrary potential, system (\ref{FOSEQU}) has the following form~\cite{KPTVV2016}:
\begin{equation}
\left\{
\begin{split}
\dot\varphi&=\psi,\\
\dot\psi&={}-3H\psi -\frac{1}{6}\left(6-K\varphi^2\right)V'+\frac{2}{3} K\varphi V,\\
\dot H&={}-\frac{K}{6}\left[2\varphi^2H^2+\left(4H\psi+V'\,\right)\varphi+2\psi^2\right].
\end{split}
\right.
\label{FOSEQUc}
\end{equation}
We integrate this system with $V=V_c$  numerically.

We consider  such a positive $\varphi_b$  that $\varphi^{+}_1<\varphi_b<\sqrt{6/K}$. The evolution of the scalar field starts at the bounce point with a negative velocity, defined by the relation
\begin{equation*}
\dot{\varphi}_b={}-\sqrt{{}-2V(\varphi_b)}.
\end{equation*}
 The field  $\varphi$  can come to zero passing the maximum of the potential. So, we keep in mind that the following subsequence of inequalities:
\begin{equation*}
0<\varphi_m<\varphi^{+}_1<\varphi_b<\sqrt{\frac{6}{K}}.
\end{equation*}

\begin{figure}[!h]
\centering
\includegraphics[width=45.2mm]{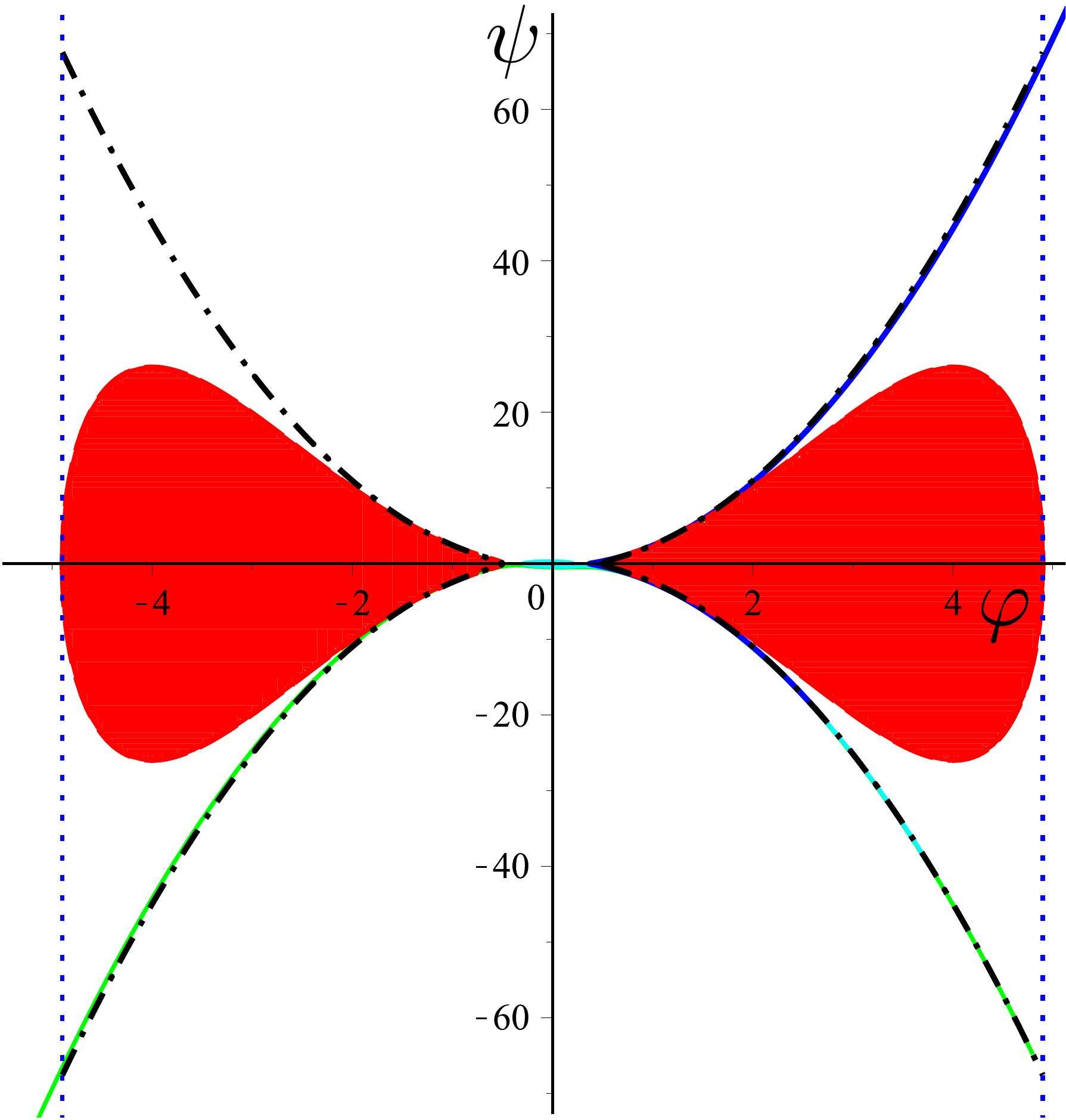}   \qquad
\includegraphics[width=45.2mm]{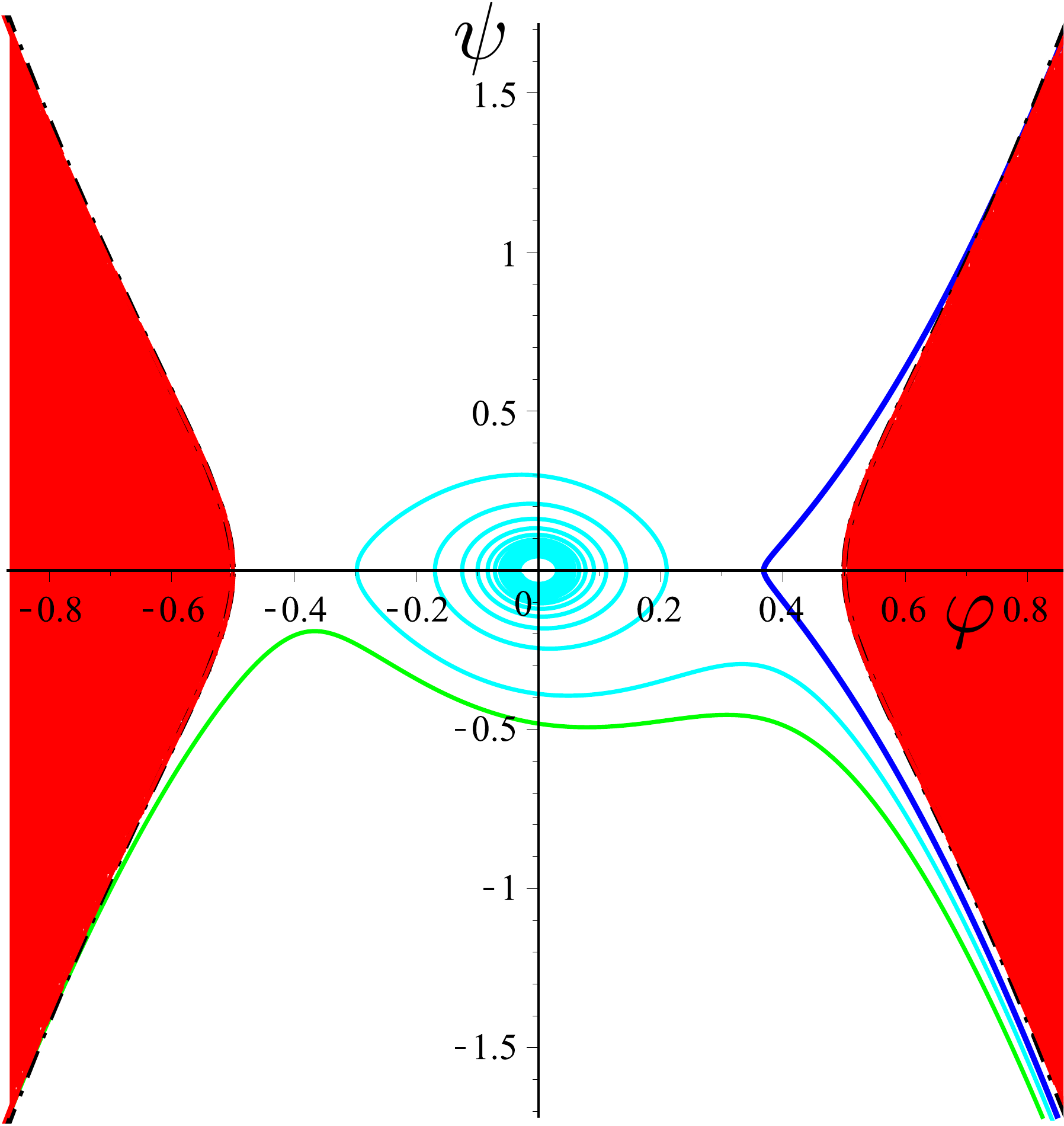}   \qquad
\includegraphics[width=45.2mm]{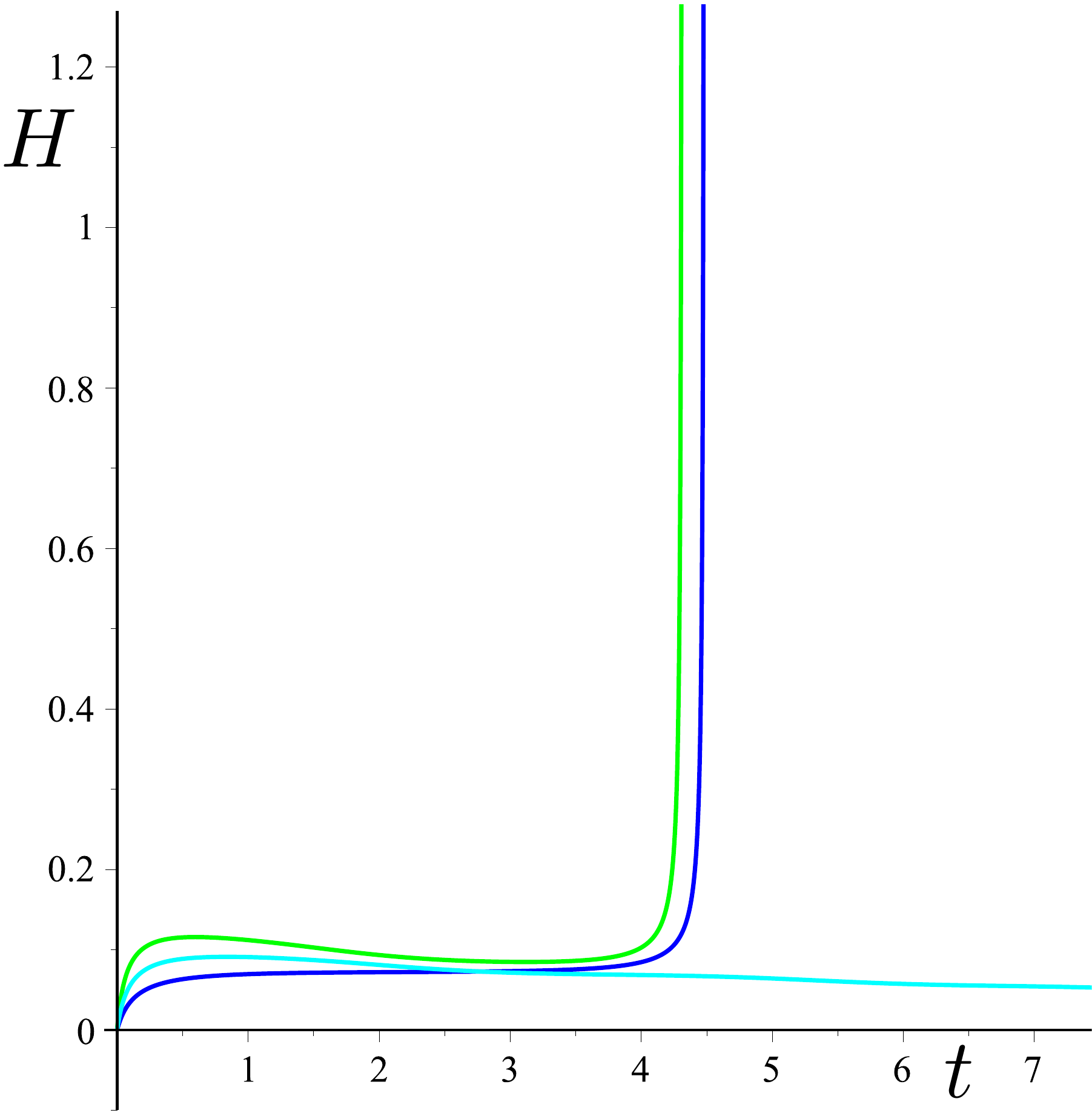}
\caption{A phase trajectory  and the Hubble parameter for the model with $U_c$ and $V_c$. The values of constants are $K=1/4$, $C_4=-4$, $C_2=1$, and $C_0=0$.
The initial conditions are $\varphi_i=2.7$, $\psi_i=-20.26259608$ (blue line), $\varphi_i=3.7$, $\psi_i=-38.36598493$ (cyan line), and $\varphi_i=4.8$, $\psi_i=-64.81244325$ (green line). The black curves are the lines of the points that correspond to $H=0$. The unreachable domain, defined by the condition $P<0$, is in red. The blue point lines correspond to $U=0$.
A zoom of the central part of phase plane is presented in the middle picture. The corresponding Hubble parameters as functions of cosmic time are presented on the right picture.}
\label{ThreeSolutions}
\end{figure}

In the case $C_0>0$ there are three possible evolutions of the bounce solutions, depending on whether the solution passes the points of the maximum of $V_{eff}$ or not. In the left and middle pictures of Fig.~\ref{ThreeSolutions} we present an example of three possible behaviors of the bounce solutions.  The solution with less initial value of $\varphi$ tends to infinity (blue curve), whereas the bounce solutions with greater initial values tends to zero (cyan curve) or to minus infinity as a  monotonically decreasing function (green curve). All trajectories start at bounce points.  The corresponding behaviors of the Hubble parameter are presented in the right picture of Fig.~\ref{ThreeSolutions} (colors the Hubble parameter evolutions on this picture coincide to the colors of the corresponding  phase trajectories).

Let us compare two bounce solutions with initial conditions: $\varphi_1(t_0)=\varphi_{b1}$ and $\varphi_2(t_0)=\varphi_{b2}>\varphi_{b1}$ and negative initial values of $\psi$.
The corresponding Hubble parameter that is equal to zero at a bounce point is given by (\ref{hpm}): $H=H_{+}$.
The initial value of $\psi$ for the first solution we denote as $\psi_{b1}$. The potential $V<0$ for all $\varphi\in [\varphi_{b1},\varphi_{b2}]$, therefore, $|\psi|\geqslant|\psi_{b1}|>0$ for any solution that pass $\varphi_{b1}$. Thus, at some moment of time $t_1>0$ the solution $\varphi_{b2}$ comes at the point  $\varphi_{b1}$: $\varphi_2(t_1)=\varphi_{b1}$.
 The value of its velocity at the bounce point $ \psi_{b1}<0$ has a minimal absolute value by comparison with  any bounce solutions that pass via $\varphi_{b1}$ with a negative velocity. It follows from (\ref{Frequoc00}),  because both $6UH^2$, and $6\dot UH=-\varphi\psi H$ are positive at $H>0$.
 It proves that if a solution with the bounce point $\varphi_{b1}$ passes through the maximum of the potential and tends to zero, then any solution with the bounce point $\varphi_{b2}>\varphi_{b1}$ tends to zero as well.

 All bounce solutions with negative initial values of $\psi$ come in the domain where $V>0$, so, $\varphi<\varphi^{+}_1$.
 If a solution does not pass the maximum of the effective potential, then after some moment $\varphi$ starts to grow and this solution comes to antigravity domain with $U_c<0 $ (the blue curves in Fig.~\ref{ThreeSolutions} denote an example of such a solution). Solutions passing the maximum of the potential can be different and need considering in detail.

If  $V<0$ for all $\varphi$, then $\varphi(t)$ is a monotonic function, because at any point $\dot\varphi\neq 0$. Similar dynamics is possible even if $V>0$ (see green curves in Fig.~\ref{ThreeSolutions}).

\begin{figure}[!h]
\centering
\includegraphics[width=45.2mm]{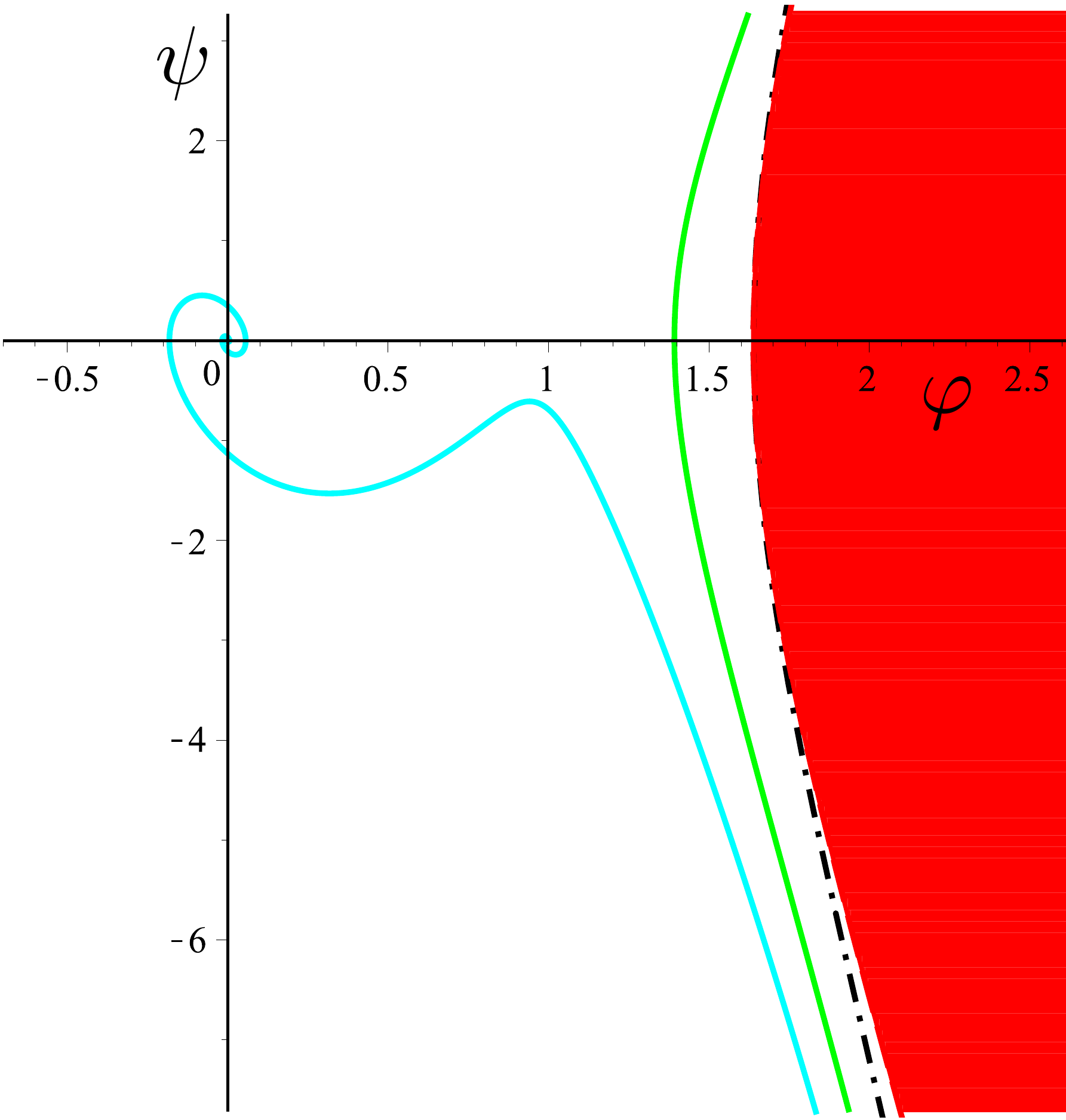}   \qquad
\includegraphics[width=45.2mm]{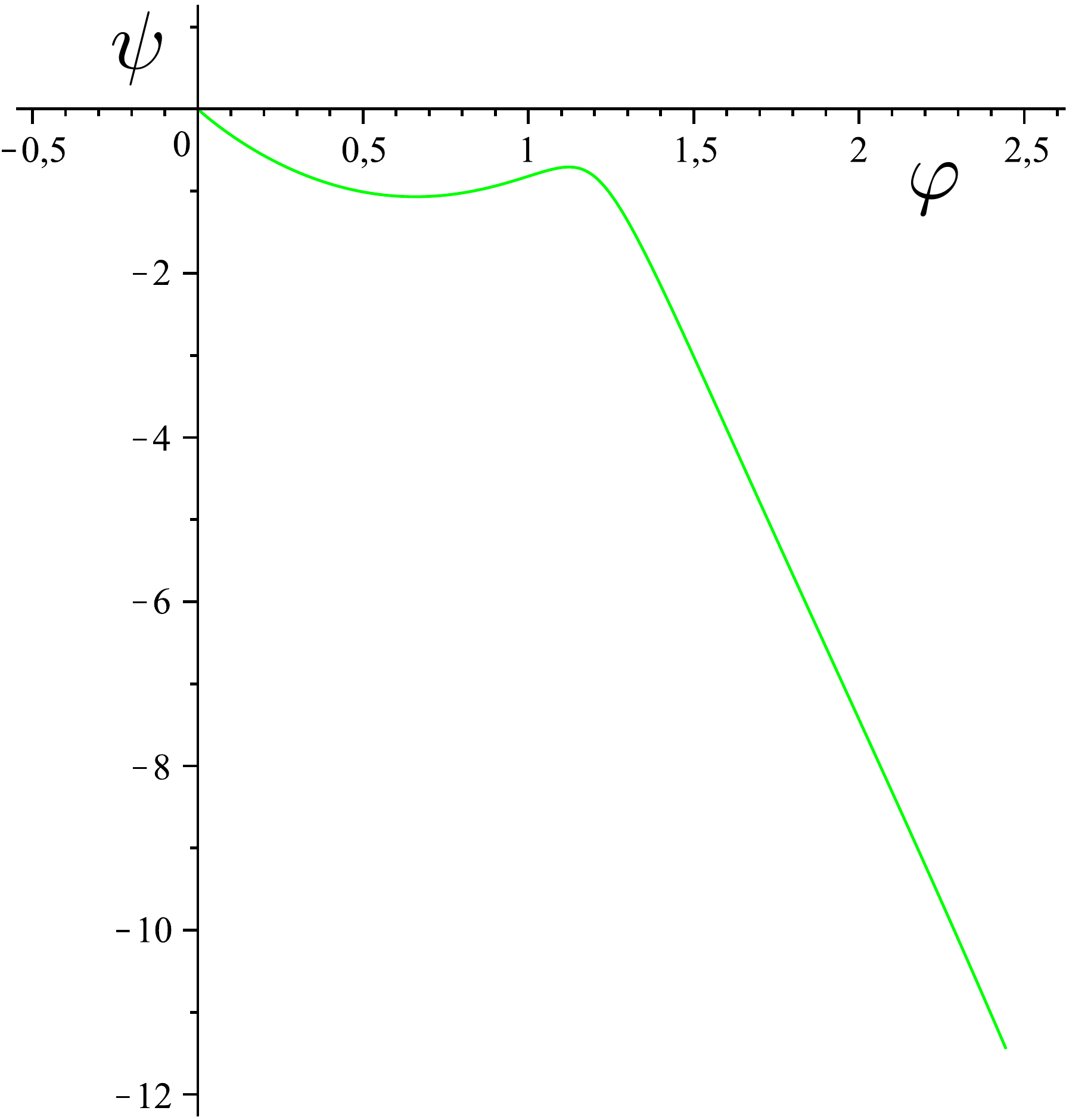}   \qquad
\includegraphics[width=45.2mm]{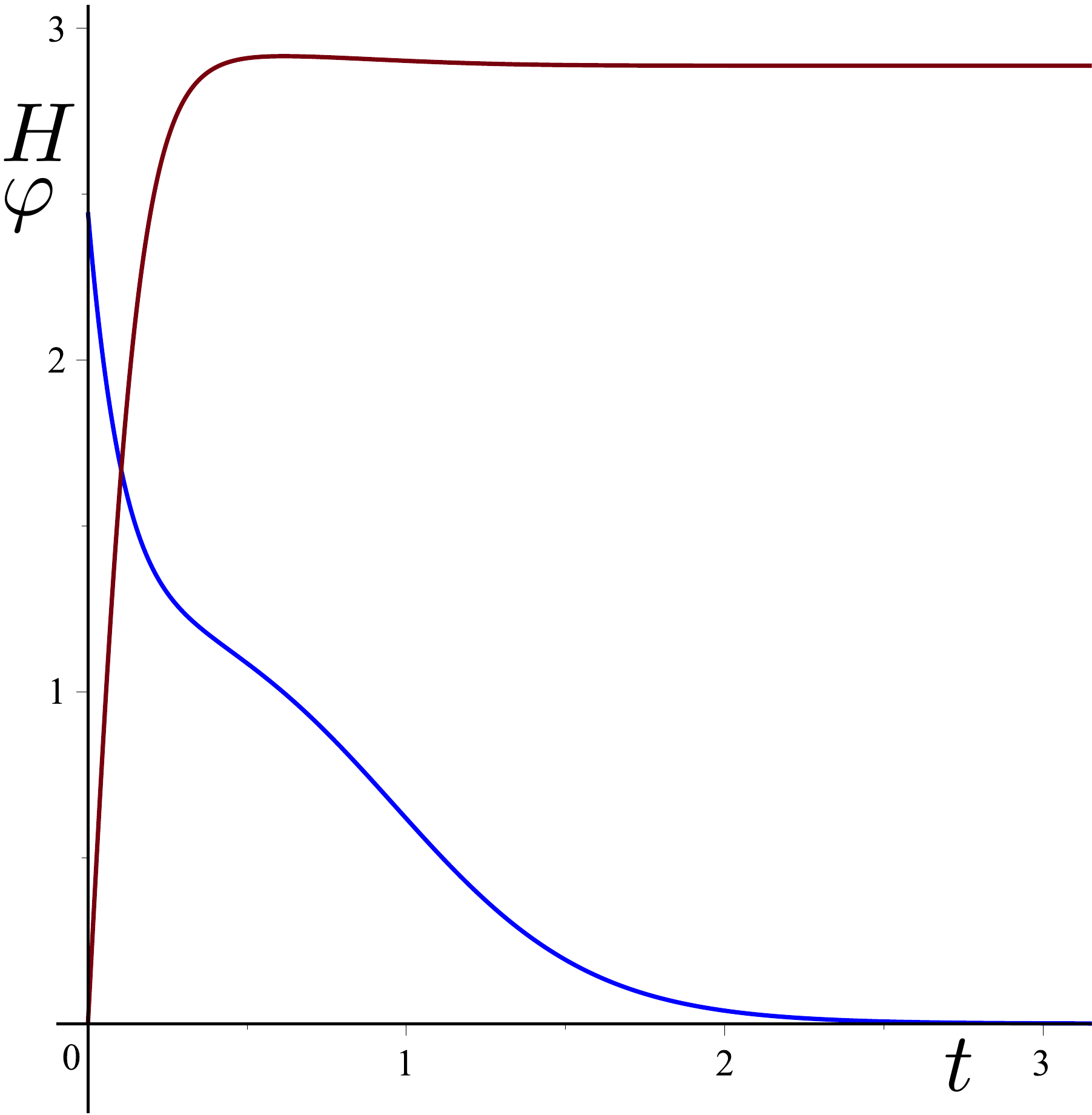}
\caption{The phase trajectory, presented in the left picture, corresponds to the following values of parameters: $K=1/4$,  $C_4=-4$, $C_2=7$ and $C_0=10$. Initial values are $\varphi_i=4.88$ and $\psi_i=-64.68078215$ (cyan curve), $\varphi_i=3.4$ and $\psi_i=-29.78638615$ (green curve). The cyan curve is an example of a stable focus. On the middle and right pictures the example of a stable node at $\varphi=0$ is presented. The field $\varphi$ (blue line) and the Hubble parameter (red line) as functions of the cosmic time are presented in the right picture.  The values of parameters are $K=1$,  $C_4=-2.7$, $C_2=1$ and $C_0=25$.
The initial conditions of the bounce solution are  $\varphi_i=2.445$, $\psi_i=-11.44650941$, and $H_i=0$.}
\label{PhaseportNode}
\end{figure}

\begin{figure}[!h]
\centering
\includegraphics[width=45.2mm]{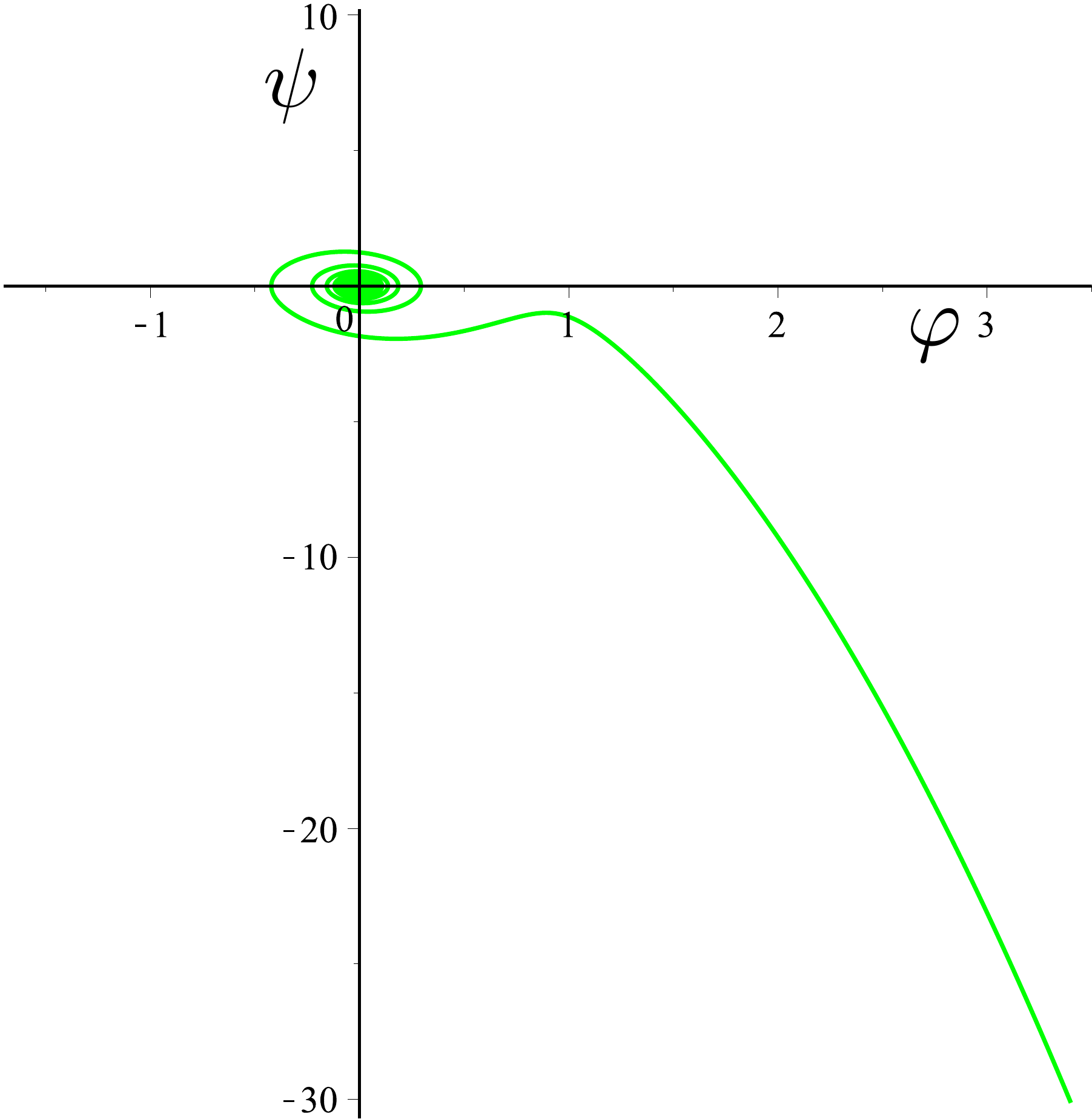} \  \  \  \  \  \  \
\includegraphics[width=45.2mm]{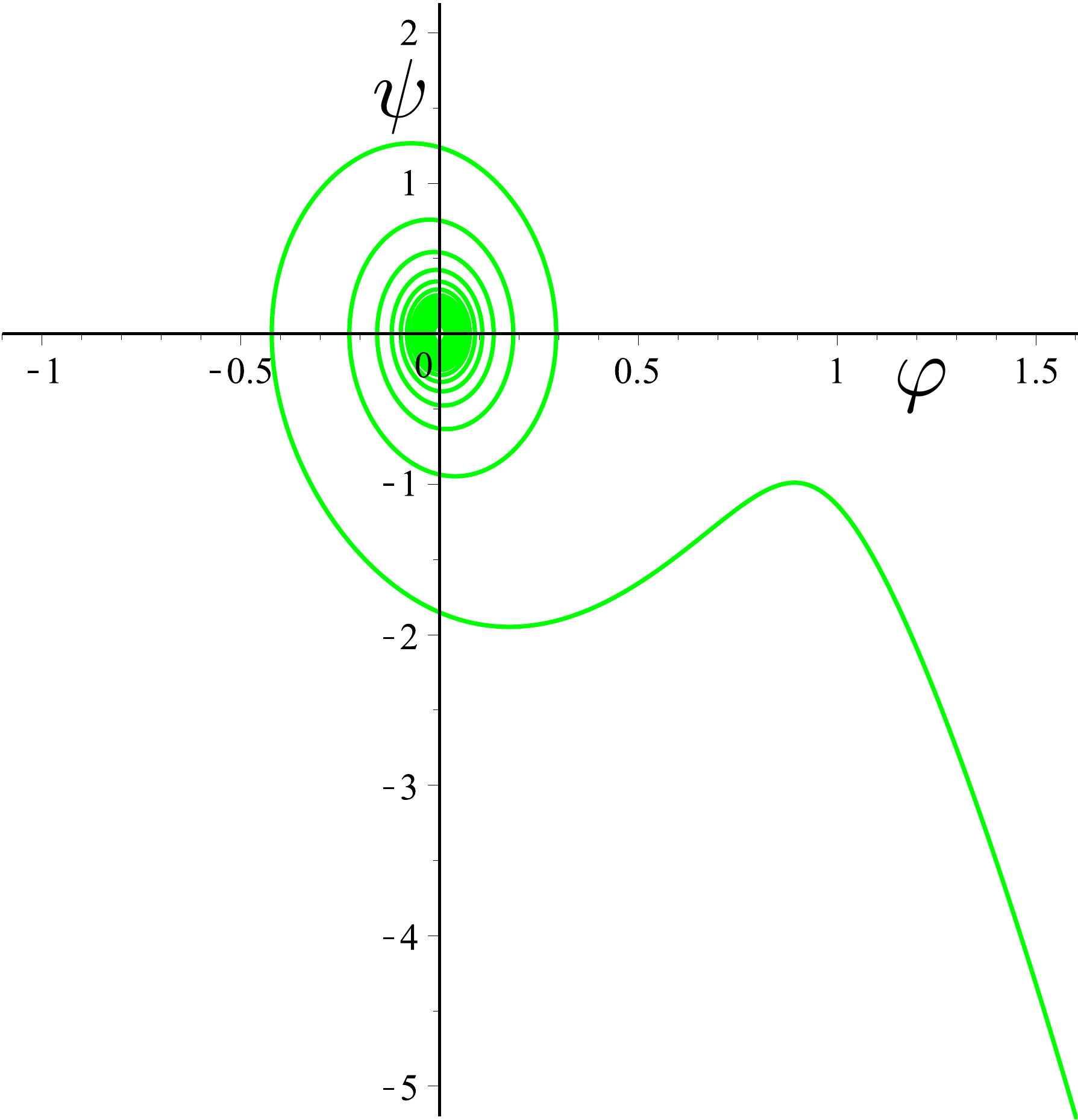} \  \  \  \  \  \  \
\includegraphics[width=45.2mm]{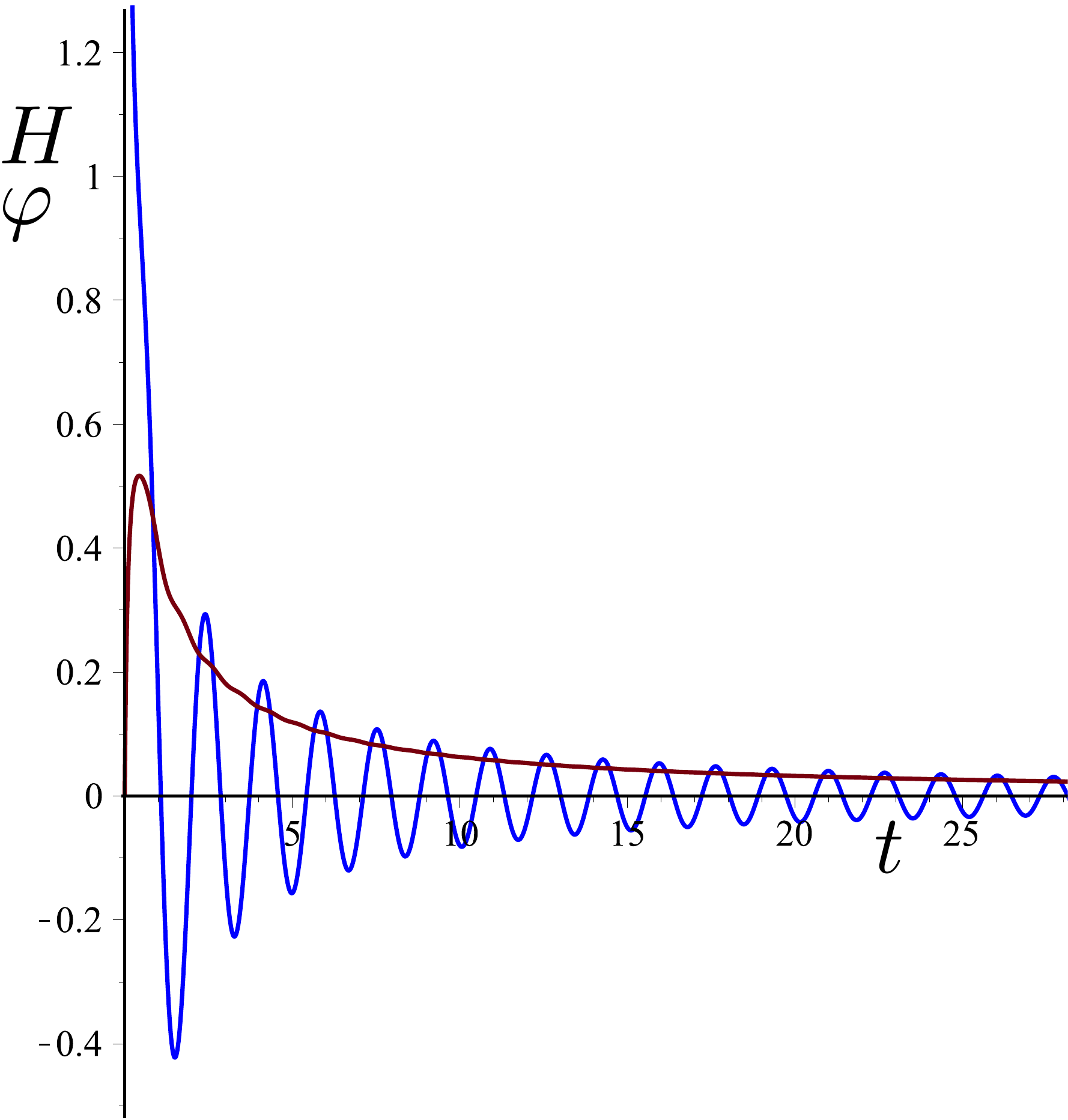}
\caption{A phase trajectory for the model with $U_c$ and $V_c$ is presented in the left picture. The values of constants are $K=1/4$, $C_4=-4$, $C_2=7$, $C_0=0$.
The initial conditions are $\varphi_i=3.4$ and $\psi_i=-30.12023904$. A zoom of the central part of the phase plane is presented in the middle picture.
The Hubble parameter (red) and the scalar field (blue) of functions of cosmic time are presented in the right picture.}
\label{PhaseportV4C0}
\end{figure}

\begin{figure}[!h]
\centering
\includegraphics[width=45.2mm]{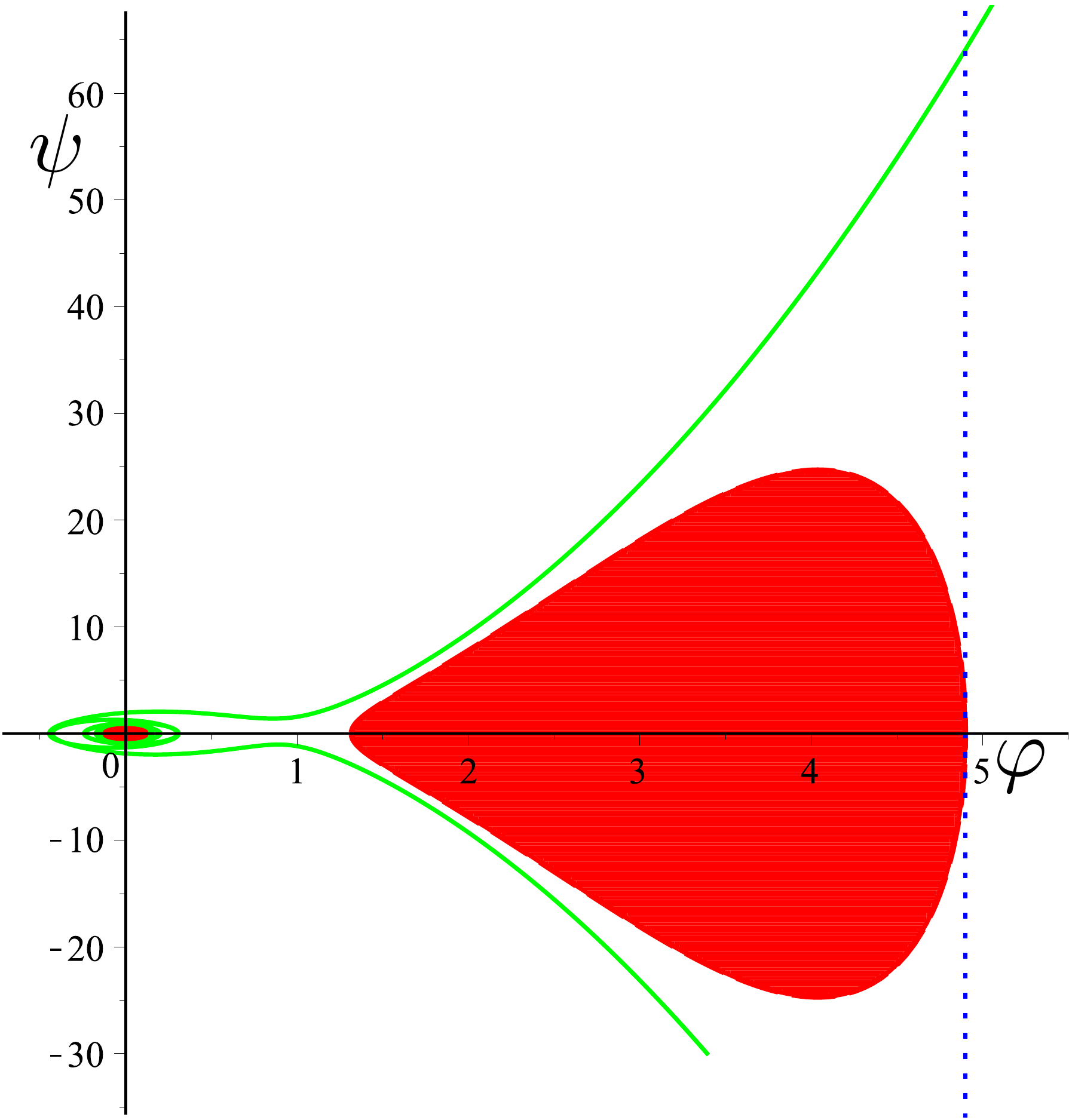}   \qquad
\includegraphics[width=45.2mm]{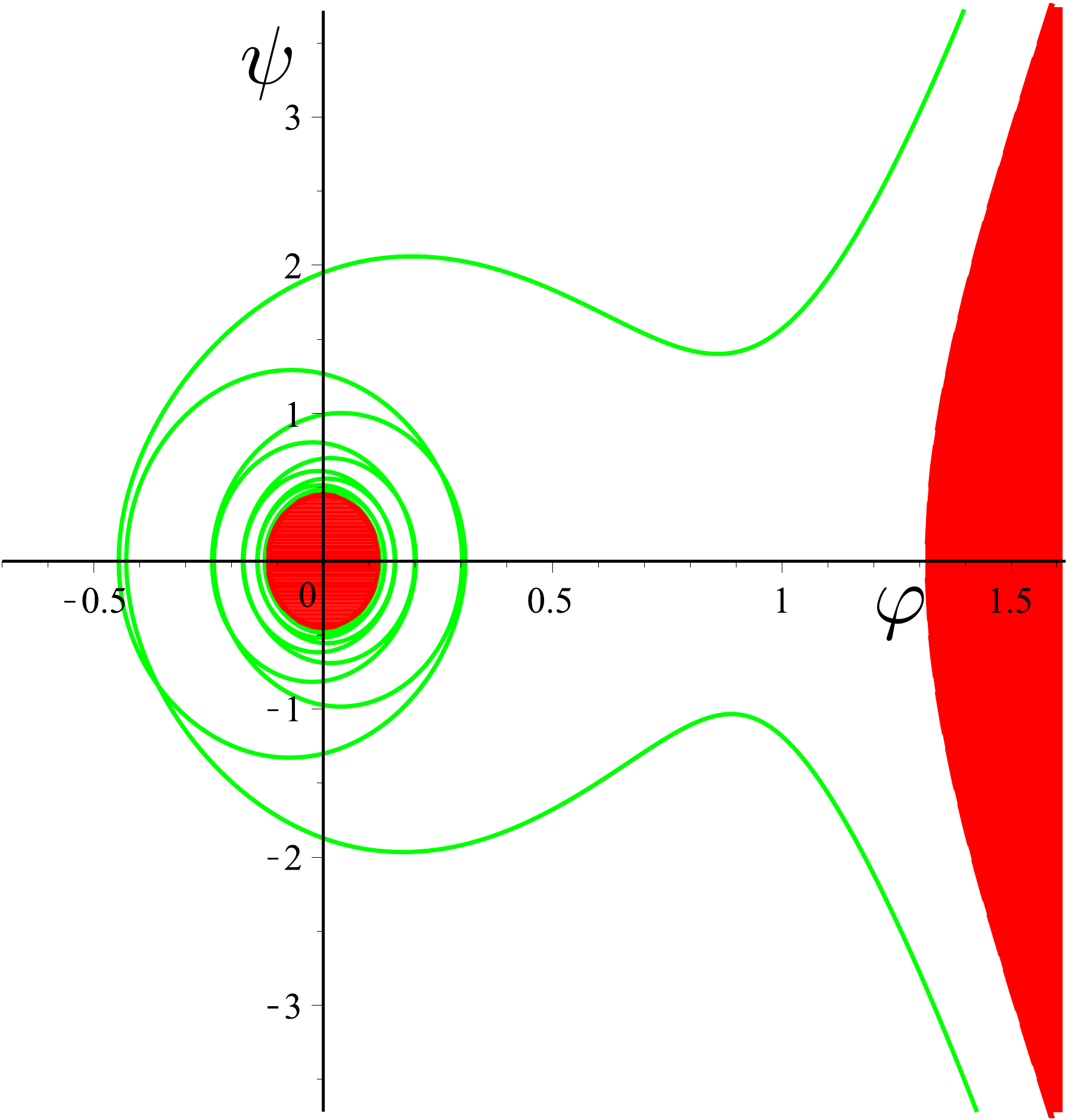}   \qquad
\includegraphics[width=45.2mm]{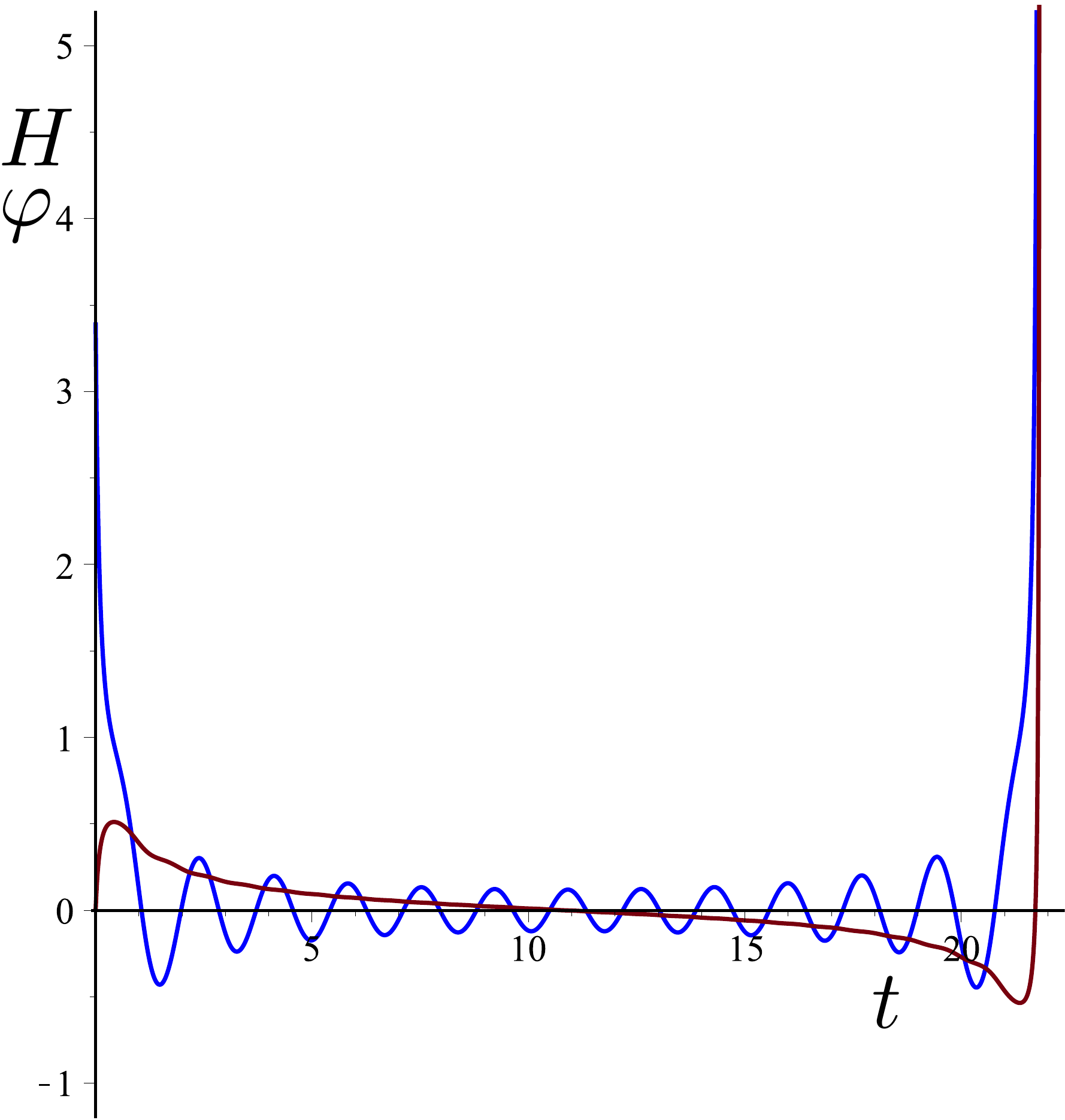}
\caption{A phase trajectory for the model with $U_c$ and $V_c$ is presented in the left picture. The values of constants are $K=1/4$, $C_4=-4$, $C_2=7$, $C_0=-0.1$.
The initial conditions are $\varphi_i=3.4$ and $\psi_i=-30.12355889$. A zoom of the central part of the phase plane is presented in the middle picture. The Hubble parameter $H(t)$ (red) and the scalar field $\varphi(t)$ (blue) are presented in the right picture.}
\label{PhaseportV4Cnega}
\end{figure}

For $C_0>0$ there exists the stable de Sitter solution $\varphi_{dS}=0$ and $H_{dS}=\sqrt{\frac{C_0K}{3}}$. It is
a stable node at  $KC_0-24 C_2\geqslant 0$ and a stable focus in the opposite case $C_0 K-24 C_2< 0$.
In  Fig.~\ref{ThreeSolutions} (cyan curves), in the left picture of Fig.~\ref{PhaseportNode}, and Fig.~\ref{PhaseportV4C0} solutions in the case of
a stable focus are presented. An example with a stable node at $\varphi=0$ is given in the middle picture of  Fig.~\ref{PhaseportNode}.
One can see that in the case of a stable node the presented Hubble parameter (the right picture of  Fig.~\ref{PhaseportNode}) is close to the Hubble parameter obtained in the paper~\cite{Boisseau:2016pfh} that has a maximum. Note that in the integrable case $C_2=0$ and  there exists a stable node at $\varphi=0$.
In the left picture of Fig.~\ref{PhaseportNode} the phase trajectories have been constructed for $K=1/4$, $C_4=-4$, $C_2=7$ and $C_0=10$. Let us now change the value of $C_0$ only and consider the model with $C_0=0$. In Fig.~\ref{PhaseportV4C0} the corresponding phase trajectory is presented. We see that now the bounce solution that starts at $\varphi_i=3.4$ tends to zero and finishes at the point $\varphi=0$. We can see that the trajectories that revolve around $(0,0)$ point look similar at  $C_0=10$ and at $C_0=0$.
Let us consider now the phase trajectory at  $C_0=-0.1$ that is presented in Fig.~\ref{PhaseportV4Cnega}. We see that trajectories are similar at the beginning only. The scalar field tends to infinity and the system comes to antigravity domain with $U_c<0$.  The form of the effective potential does not depend essentially from the sign value of $C_0$ (see Fig.~\ref{VeffC0}), but the sign of $V_{eff}(0)=C_0$ is different. By this reason, the behavior of bounce solutions are essentially different. In the right pictures of Fig.~\ref{PhaseportV4C0} and
Fig.~\ref{PhaseportV4Cnega} one  can see that the behavior of the Hubble parameter also essentially depends on the sign of~$C_0$.

\begin{figure}[!h]
\centering
\includegraphics[width=72mm]{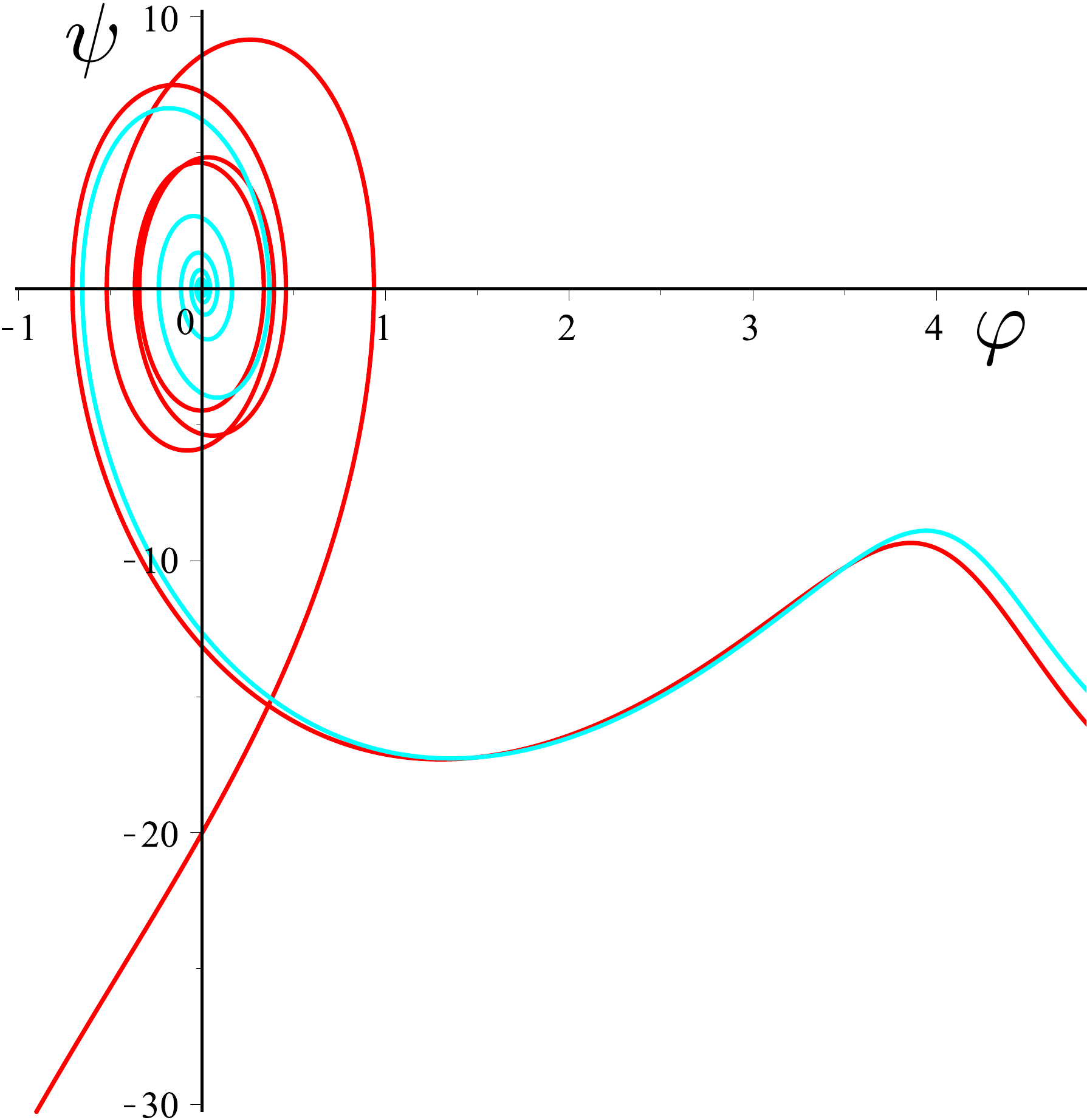}     \qquad
\includegraphics[width=72mm]{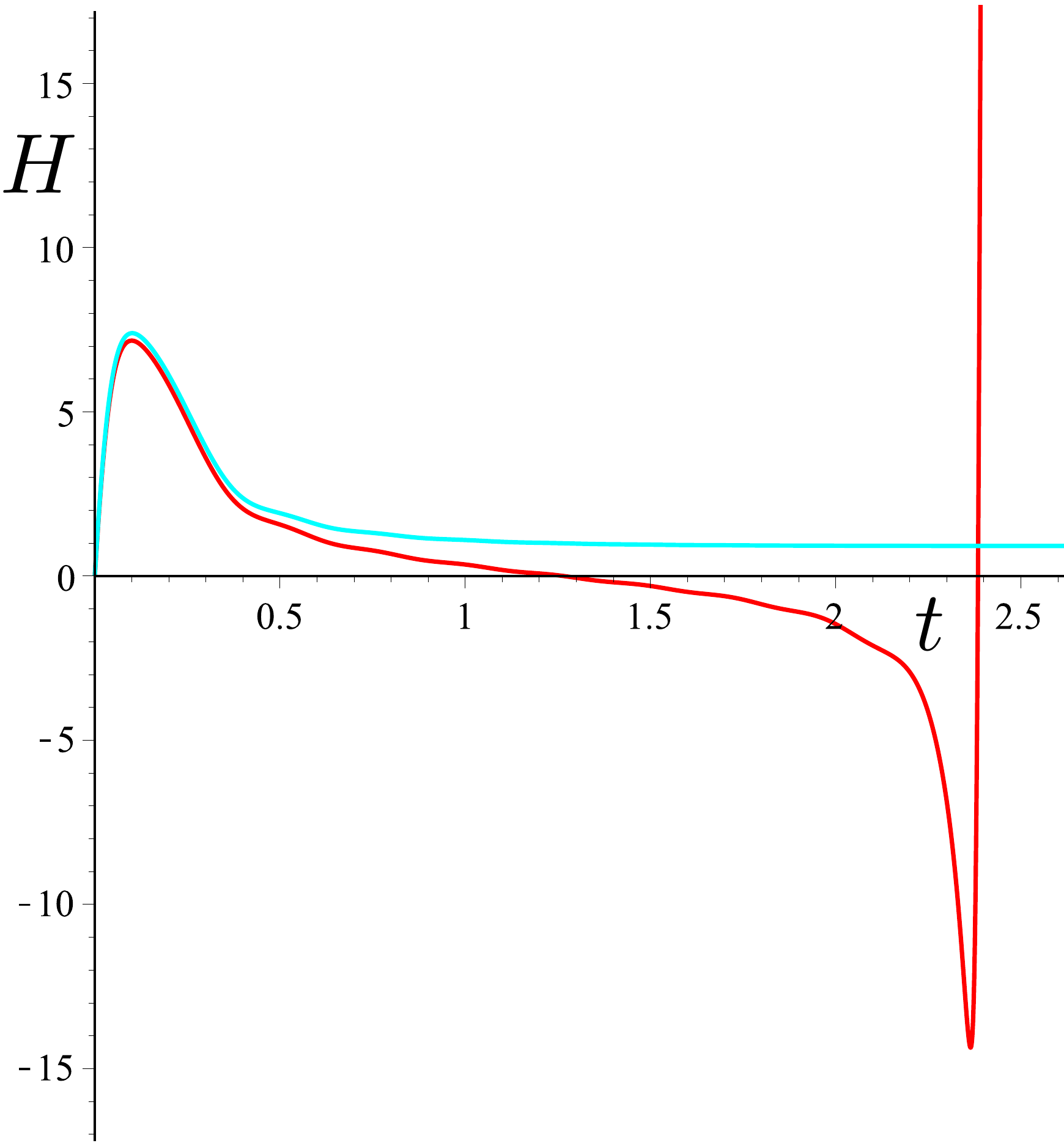}
\caption{The phase trajectories (right picture) and the corresponding Hubble parameters (left picture) are presented.  The values of parameters are $K=1/4$,  $C_4=-4$, $C_2=90$. The parameter $C_0=10$ for cyan curves and $C_0=-10$ for red curves.
The initial conditions of bounce solution are  $\varphi_i=4.88$, $\psi_i=-15.17936692$ (cyan curve) and $\psi_i=-16.44424459$ (red curve).}
\label{C0pm}
\end{figure}

The difference between the solutions of system (\ref{FOSEQUc}) with a positive and a negative $C_0$ is demonstrated in Fig.~\ref{C0pm} as well. The cyan curves correspond to $C_0=10$, whereas the red curves correspond to $C_0=-10$. We see that the phase trajectories of the field $\varphi$ and behaviours of the Hubble parameter are similar in the beginning, but stand essentially different in the future.

We come to conclusion that the behavior of solutions essentially depends on the sign of $C_0$. To understand the reason of this dependence let us consider the domain $|\varphi|<\sqrt{6/K}$, where $U>0$. From (\ref{hpm}) it follows that the Hubble parameter is real  if
 \begin{equation}
 \label{conddotsigma}
 \dot{\varphi}^2\geqslant{}-\frac{2UV}{U+3{U'}^2}={}-4KUV.
\end{equation}

If the constants $C_i$ are such that $V\geqslant0$ for all  $\varphi \in [-\varphi_m,\varphi_m]$, then this condition is always satisfied, and  the  field  $\varphi$ tends to a minimum of $V_{eff}$  at $\varphi=0$.
We see such evolutions  in Fig.~\ref{ThreeSolutions}, Fig.~\ref{PhaseportNode}, and Fig.~\ref{PhaseportV4C0}. Note that in this case if $P>0$ in the moment when potential stands positive and the function $\varphi$ tends to zero, then $P>0$ at any  moment in future.

If the constants $C_i$ are such that the potential change the sign and $V(0)<0$, then the evolution is different (see Fig.~\ref{PhaseportV4Cnega}).  The Hubble parameter becomes negative and positive again, so, there are two bounce points. After that the Hubble parameter tends to infinity.

Let us consider this case in detail. If $C_0<0$, then there is a restricted domain in the neighborhood of $(0,0)$ point on phase plane such that the values of the scalar field $\varphi$ and its derivative correspond to non-real values of the Hubble parameter. The boundary of this domain is defines by equation $P=0$. In Fig.~\ref{PhaseportV4Cnega} we see that the phase trajectory rotates around this domain. The trajectory can not cross the boundary, but can touch it.

We show that all such trajectories touch the boundary at some finite moment of time.  Let for some moments of time $t_1$ and $t_2>t_1$ we have $\varphi(t_2)=\varphi(t_1)$, then, using $U+3{U'}^2=\frac{1}{2K}$ and formula (\ref{Fr21Qm}), we get
\begin{equation*}
P(t_2)-P(t_1)={}
-\int\limits_{t_1}^{t_2}\frac{U+3{U'}^2}{4 U\sqrt{U}}\psi^2\,dt=-\int\limits_{t_1}^{t_2}\frac{1}{8 KU\sqrt{U}}\psi^2\,dt\leqslant \tilde{C}<0.
\end{equation*}
where $\tilde{C}$ is a negative number. Therefore, this integral has a finite negative value. For any circle value of $P$ decreases on some positive value, which doesn't tend to zero during evolution, when number of circles increase. We come to conclusion that only a finite number of circles is necessary to get the value $P=0$. At this point $\dot P<0$ as well, so the function $P$ changes the sign. When $P=0$ two possible values of the Hubble parameters: $H_+$ and $H_-$ coincide. At this moment the value of the Hubble parameter changes from $H_+$ to $H_-$. The value of the function $P$ continues to decrease, so, the distance between the trajectory and the boundary of unreachable domain increases. We do not say that its increase monotonically but the absolute value of $P$ that is a characteristic of this distance increases on a finite quantity after any circle. So, after some finite number of circles the absolute value of $\varphi$ becomes more then $\varphi_m$. After this moment   $\varphi$  monotonically tends to infinity and at some finite moment we get $\varphi^2=6/K$. Thus, the final of trajectory is in the antigravity domain always.

Note that in the domain with $V>0$ the Hubble parameter is uniquely defined as a function $\varphi$ and $\psi$ by~(\ref{hpm}). If $C_0\geqslant 0$, then
 whole evolution of bounce solutions is evolution a solution of the second order system, whereas for $C_0< 0$ the third order system~(\ref{FOSEQUc}) with the additional condition~(\ref{Frequoc00}) is not equivalent to any  second order system.

\section{Models with monomial potentials and cosmological constant}
\subsection{Conditions of the bounce existence}
In this section we consider the scalar field potential of the form
\begin{equation}
\label{V2}
V(\varphi)=C_n\varphi^n+C_0,
\end{equation}
where $n$ is an even natural number, and following non-minimally coupled function
\begin{equation}
U(\varphi)=\frac{1}{2K}-\frac{\xi\varphi^2}{2},
\end{equation}
where $\xi$ is positive.

We are interested in cosmological scenarios in the physical region $G_{eff}>0$, where the bounce occurs at first and after that stable de Sitter solution $H_{dS}=\sqrt{\frac{V(\varphi_{dS})}{6U(\varphi_{dS})}}$  is realized. So, the potential should change the sign. The potential $V_{eff}$ is an even function,
hence, it has an extremum at $\varphi=0$. If we additionally suppose that $\varphi_{dS}=0$, then we get the condition $C_0>0$. Therefore, the bounce condition $H(\varphi_b)=0$ for the chosen potential $V$ gives:
\begin{equation}
\label{bounce1}
V(\varphi_b)\leqslant 0~~\Rightarrow ~~C_n{\varphi_b}^n+C_0\leqslant 0~~\Rightarrow ~~C_n<0.
\end{equation}

From the condition $\dot H(\varphi_b)>0$ it follows (see \eqref{bounceCond2}):
\begin{equation}
\label{bounce2}
V(\varphi_b)(1-2\xi)>\xi\varphi_b V'(\varphi_b)~~\Rightarrow ~~C_n{\varphi_b}^n(1-\xi(2+n))>(2\xi-1)C_0.
\end{equation}

  Taking into account the inequalities  $C_n<0$ and $C_0>0$, we obtain from conditions~\eqref{bounce1} and \eqref{bounce2} that the bounce exists for
\begin{itemize}
\item(i) $0<\xi<\frac{1}{n+2}$
\begin{itemize}
\item(1). If  $\displaystyle{}-\frac{C_0}{C_n}\frac{1-2\xi}{1-\xi(2+n)}<\displaystyle{}{\left( \frac{1}{K\xi}\right) }^{\frac{n}{2}}$,~~ then ~~$\displaystyle{}-\frac{C_0}{C_n}\leqslant{\varphi_b}^n<\displaystyle{}-\frac{C_0(1-2\xi)}{C_n(1-\xi(2+n))}$.
\item(2). If $-\frac{C_0}{C_n}\frac{1-2\xi}{1-\xi(2+n)}\geqslant{\left( \frac{1}{K\xi}\right) }^{\frac{n}{2}}$,~~ then ~~$-\frac{C_0}{C_n}\leqslant{\varphi_b}^n<{\left( \frac{1}{K\xi}\right) }^{\frac{n}{2}}$.
\end{itemize}
\item(ii) $\xi\geqslant\frac{1}{n+2}$
\\$-\frac{C_0}{C_n}\leqslant{\varphi_b}^n<{\left( \frac{1}{K\xi}\right) }^{\frac{n}{2}}$.
\end{itemize}
Here $\varphi_b$ is the value of the scalar field at the bounce. We see that there is an additional restriction on the location of points
of bounce in comparison with the cases studied earlier in~\cite{Boisseau:2015hqa,Boisseau:2016pfh}, appearing when
 $0<\xi<\frac{1}{n+2}$ (the case (i1)). This happens due to $\dot H>0$ condition of the bounce, which can be violated in the case of small enough $\xi$, causing a recollapse instead of bounce.

The bounce point should corresponds to $G_{eff}>0$ that gives the following condition
\begin{equation}
\label{phicrit}
\varphi_b^2<\frac{1}{K\xi}.
\end{equation}

The effective potential $V_{eff}(\varphi)$ has zeros for
\begin{equation}
\label{zerosVn}
{(\varphi_{zero})}^n={}-\frac{C_0}{C_n}.
\end{equation}
They are located in the region $G_{eff}>0$ only if
\begin{equation}
\label{xicrit}
{(\varphi_{zero})}^2={\left( -\frac{C_0}{C_n}\right)}^{\frac{2}{n}} <\frac{1}{K\xi}\Rightarrow
\xi<\xi_{cr}=\frac{1}{K}{\left( -\frac{C_n}{C_0}\right)}^{\frac{2}{n}}.
\end{equation}
Thus, we have received a restriction from above for $\xi$.

The first derivative of $V_{eff}$ with respect to $\varphi$ for the chosen functions $V$ and $U$ is given~by
\begin{equation}
\label{dVeff}
V'_{eff}=\frac{\varphi(C_n K\xi\varphi^n(4-n)+n C_n\varphi^{n-2}+4\xi K C_0)}{{(1-K\xi\varphi^2)}^3}.
\end{equation}
Therefore, the effective potential has the extremum at the point $\varphi=0$ for any even $n$. Non-zero extrema $\varphi_m$ for $n=2$
are
\begin{equation}
\label{extrn2}
\varphi_m=\pm\sqrt{-\frac{C_2+2\xi K C_0}{C_2 K\xi}}.
\end{equation}
For $n=6$ values of $\varphi_m^2$ are roots of the following cubic equation
\begin{equation}
\label{extrn6}
{\left(\varphi_m^2\right)}^3-\frac{3}{K\xi}{\left(\varphi_m^2\right)}^2-\frac{2C_0}{C_6}=0,
\end{equation}
which has three real roots for $\xi^3<-\frac{2C_6}{K^3C_0}$. We consider only $\xi^3<\xi_{cr}^3=-\frac{C_6}{K^3C_0}<-\frac{2C_6}{K^3C_0}$ (see \eqref{xicrit}), hence, there are three real roots in this case. Only positive ones have the physical sense ($\varphi_m^2>0$).

Let us study properties of the effective potential $V_{eff}$ to analyse the Lyapunov stability of de Sitter solutions.
We calculate $A(\varphi_{dS})=K^2>0$ for de Sitter solution $\varphi_{dS}=0 $, \  $H_{dS}=\sqrt{\frac{C_0 K}{3}}$. Then it is stable for $V''_{eff}(0)>0$, namely,
\begin{equation}
\begin{array}{l}
C_2+2K\xi C_0>0, ~\quad n=2,\\
C_0>0, ~\quad n>2.
\end{array}
\end{equation}
From this it follows that in the case of $n=2$ there is the restriction from below for the coupling constant
 \begin{equation*}
\xi>-\frac{C_2}{2KC_0}.
\end{equation*}
 On the other hand, for $n>2$ the interval of possible $\xi$ allowing evolution towards de Sitter solution after the bounce is not restricted from below (we remind that $K>0$ and we consider only positive values of $\xi$ in this paper).

When de Sitter solution is stable we have (see \eqref{node} and \eqref{focus}) a monotonic decreasing of the scalar field for
\begin{equation}
\label{noden}
\begin{array}{l}
\xi\leqslant\frac{3}{16}-\frac{C_2}{2K C_0}, ~~n=2\\
\\\xi\leqslant\frac{3}{16}, ~~n>2
\end{array}
\end{equation}
or scalar field oscillations for
\begin{equation}
\label{focusn}
\begin{array}{l}
\xi>\frac{3}{16}-\frac{C_2}{2K C_0}, ~~n=2\\
\\\xi>\frac{3}{16}, ~~n>2.
\end{array}
\end{equation}
This result is in agreement with the one received in~\cite{Boisseau:2016pfh} for $n=4$.

Let us summarize our requirements to the effective potential $V_{eff}(\varphi)=\frac{C_n\varphi^n+C_0}{{(1-K\xi\varphi^2)}^2}$ and the parameter $\xi$:
\begin{itemize}
\item ~~$C_0>0$, ~~~~$C_n<0$,
\item ~~$0<\xi<\frac{1}{K}{\left({} -\frac{C_n}{C_0}\right)}^{\frac{2}{n}}$~~ for ~~$n>2$,
\item ~~${}-\frac{C_2}{2KC_0}<\xi<{}-\frac{C_2}{KC_0}$~~ for ~~$n=2$.
\end{itemize}

\subsection{Numerical investigations}

\begin{figure}[!h]
\centering
\includegraphics[width=47mm]{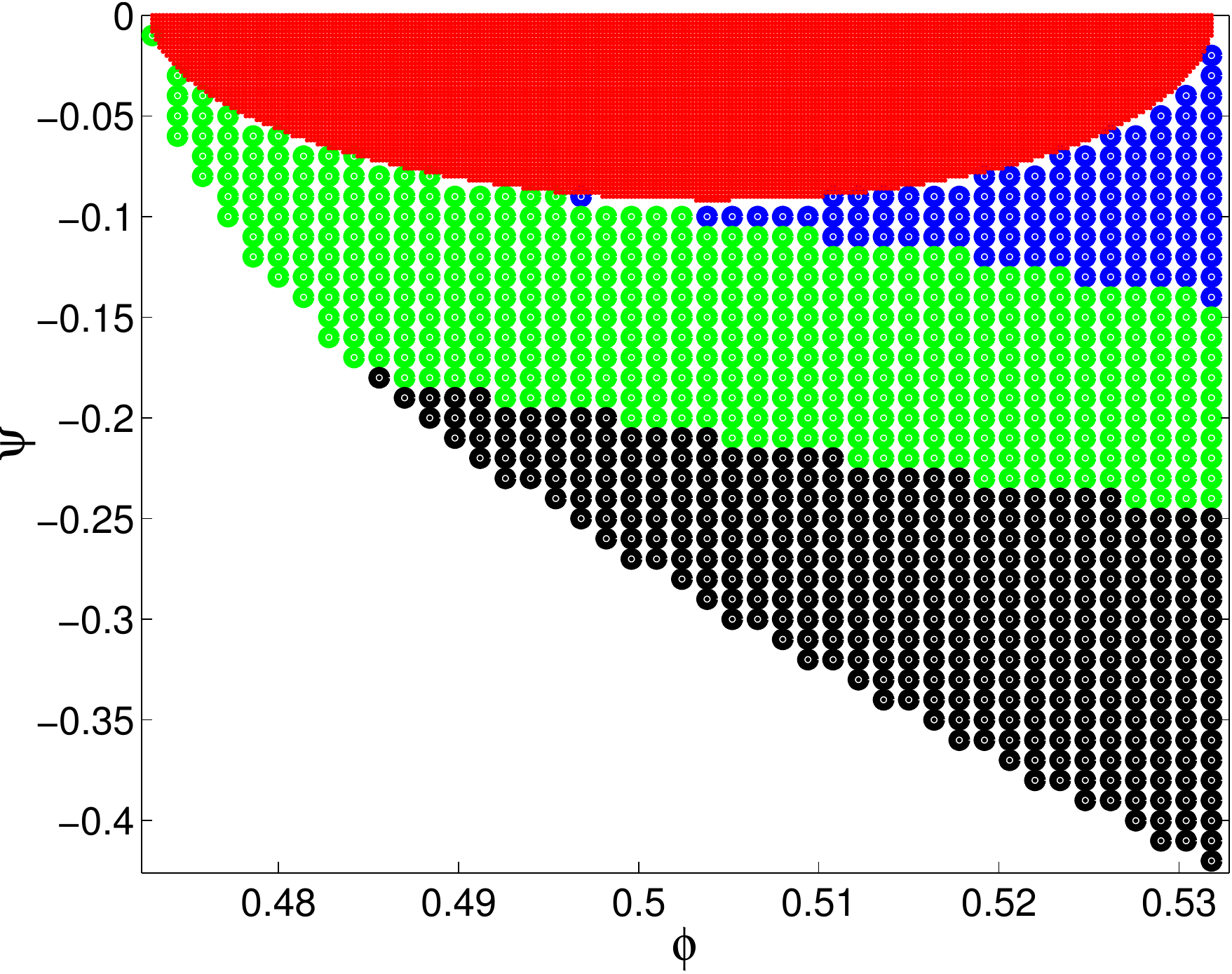}\quad
\includegraphics[width=45mm]{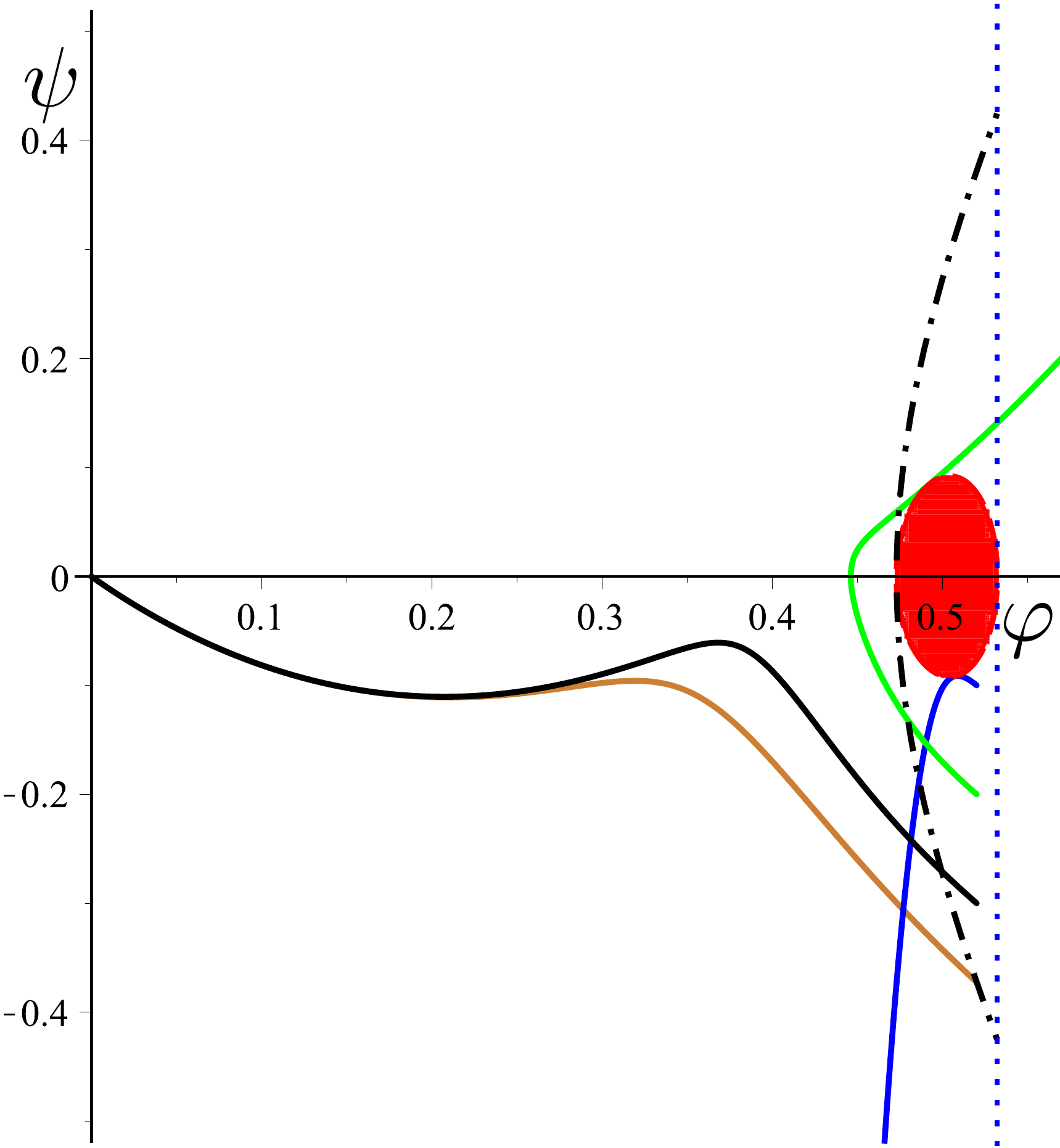}\qquad
\includegraphics[width=45mm]{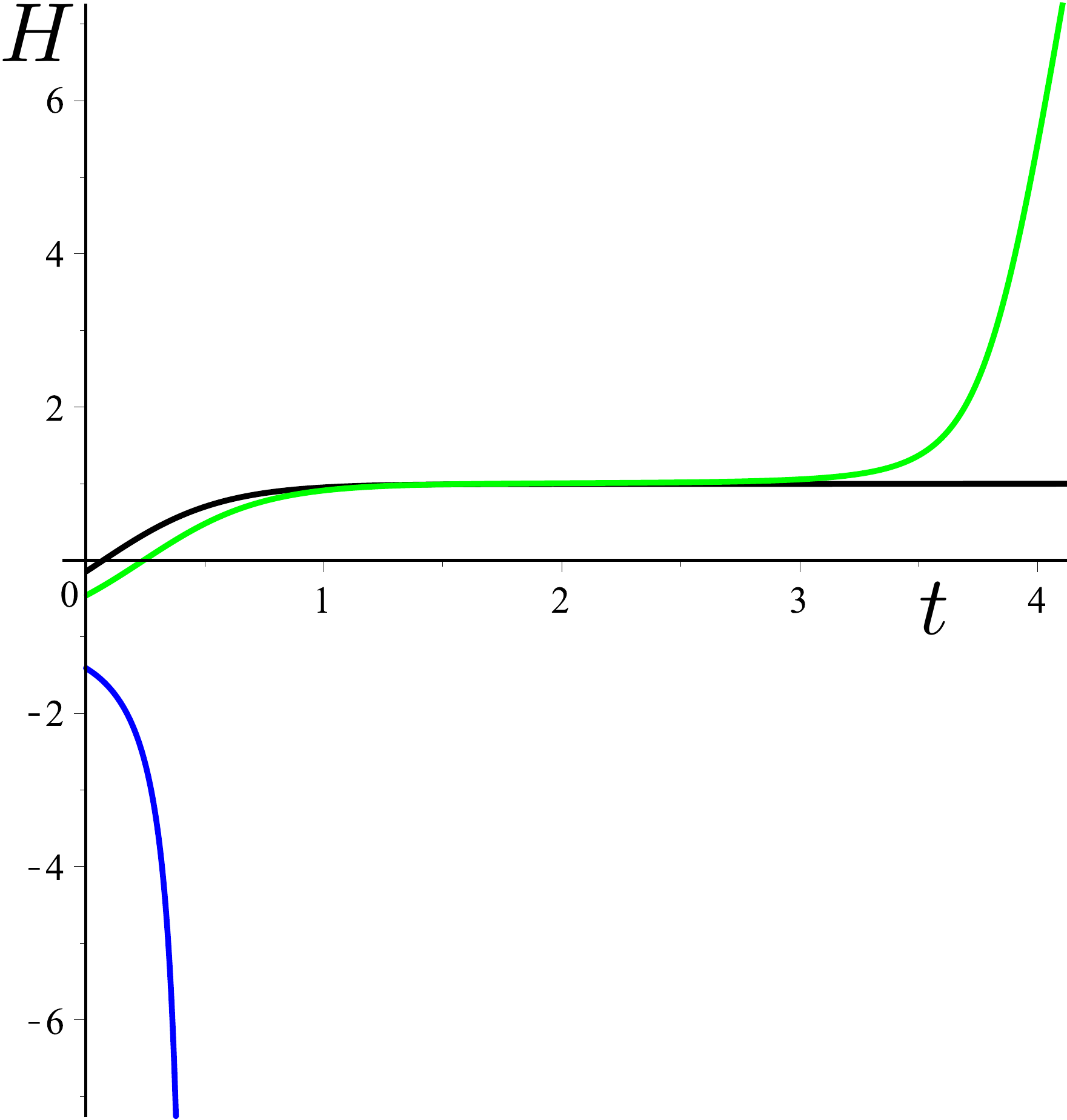}
\caption{There are basins of attraction for three possible attractors for the case of $V=C_4\varphi^4+C_0$, $U=U_0-\xi\varphi^2/2$ in the left picture: de Sitter solutions (black circles), trajectories with a bounce for $G_{eff}>0$ and then going to the antigravity region (green circles) and trajectories without a bounce for $G_{eff}>0$ going to the antigravity domain (blue circles). The parameters are $\xi=1/6+0.01$, $K=20$, $C_4=-3$, and $C_0=0.15$. Three phase trajectories (middle picture) and dependencies $H(t)$ (right picture) corresponding these attractors are plotted for initial data $\varphi_i=0.52$ and $\psi_i=-0.1$ (blue curve), $\varphi_i=-0.2$ (green curve), $\psi_i=-0.3$ (black curve). The gold trajectory starts at the bounce point $\psi_i=-0.3724204076$. The black dash line corresponds to $H=0$. The blue point line corresponds to $U=0$. The red color in the left and middle pictures indicates the unreachable domain.}
\label{n4xi01767}
\end{figure}

\begin{figure}[!h]
\centering
\includegraphics[width=72mm]{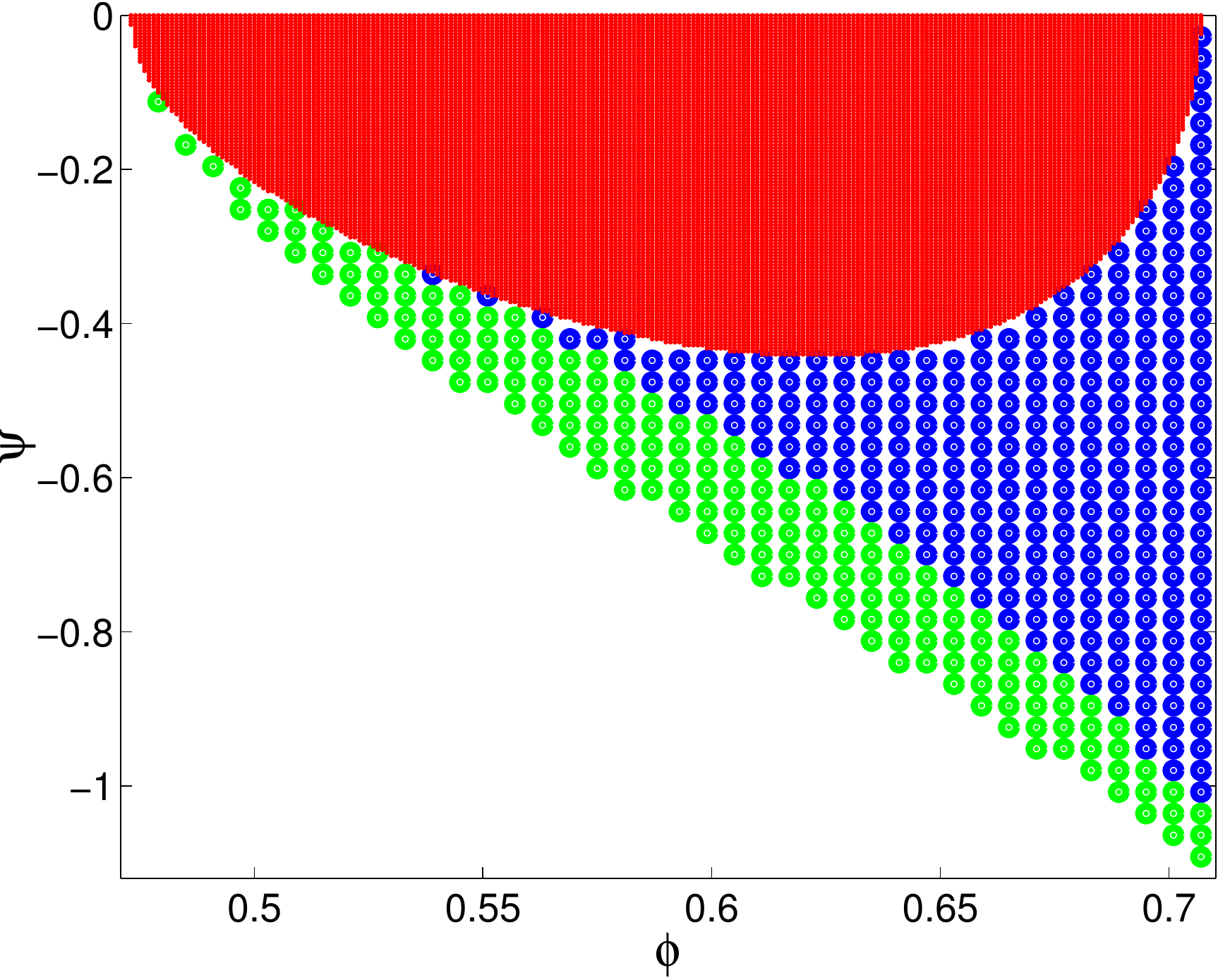} \qquad
\includegraphics[width=72mm]{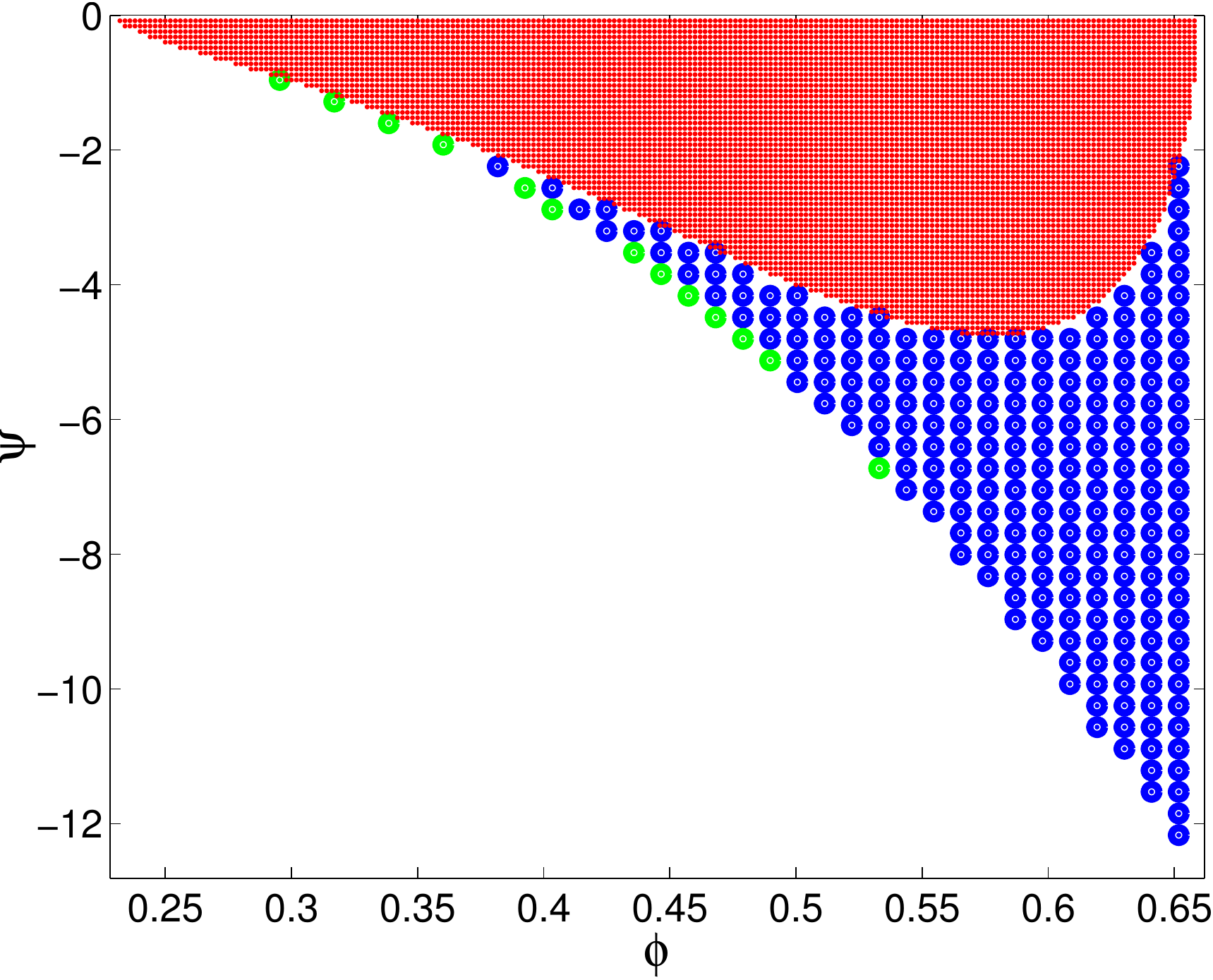}
\caption{There are basins of attraction for cases of $V=C_4\varphi^4+C_0$,  $C_4=-3$, $U=U_0-\xi\varphi^2/2$, $\xi=1/10$ (left) and $V=C_6\varphi^6+C_0$, $C_4=-1000$, $U=U_0-\xi\varphi^2$, $\xi=0.115$ (right): trajectories with a bounce for $G_{eff}>0$ and then going to the antigravity region (green circles) and trajectories without a bounce for $G_{eff}>0$ going to the antigravity domain (blue circles). The parameters are $K=20$, $C_0=0.15$. The red color corresponds to unreachable domain.}
\label{n4xi01}
\end{figure}

\begin{figure}[!h]
\centering
\includegraphics[width=47mm]{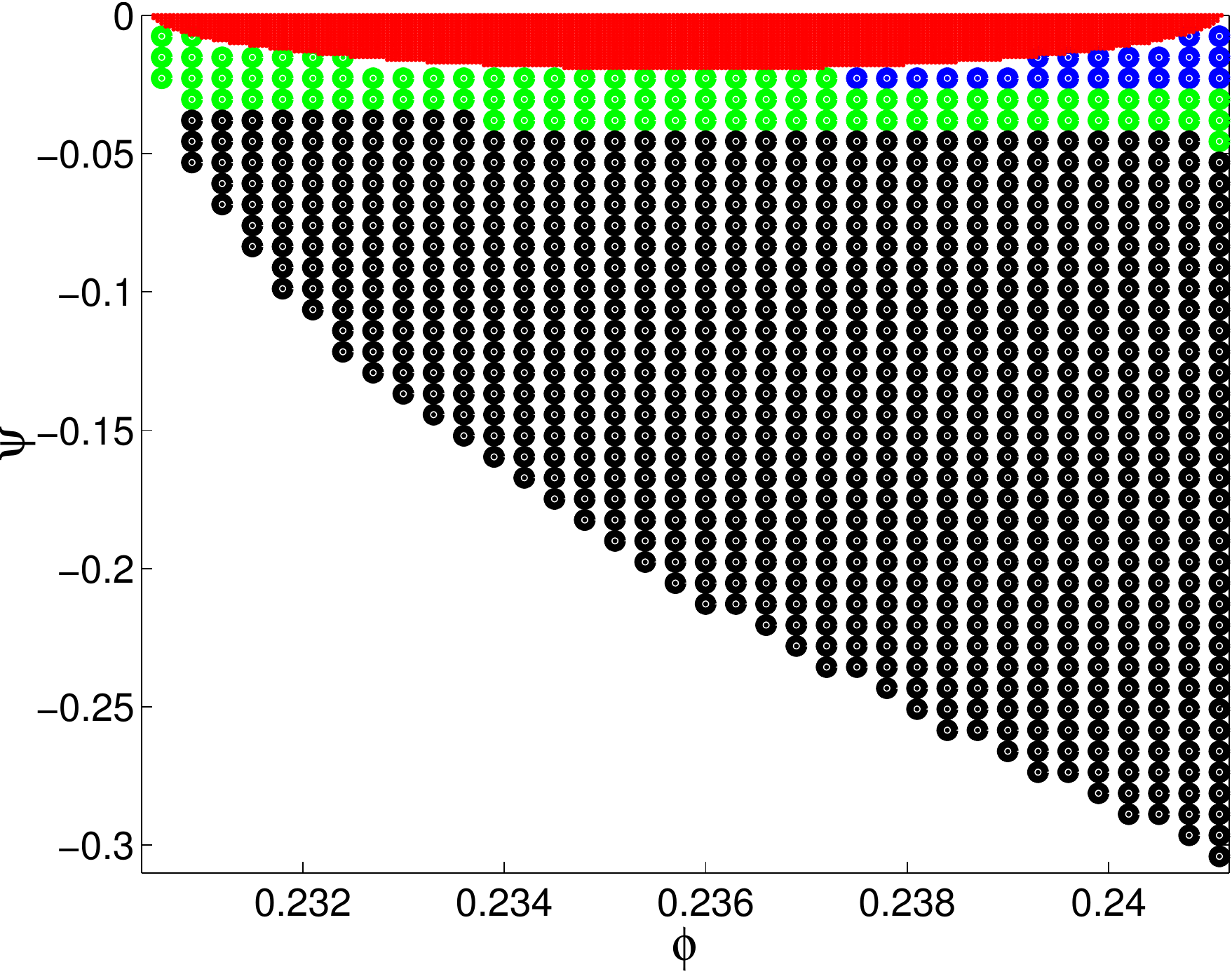} \quad
\includegraphics[width=45mm]{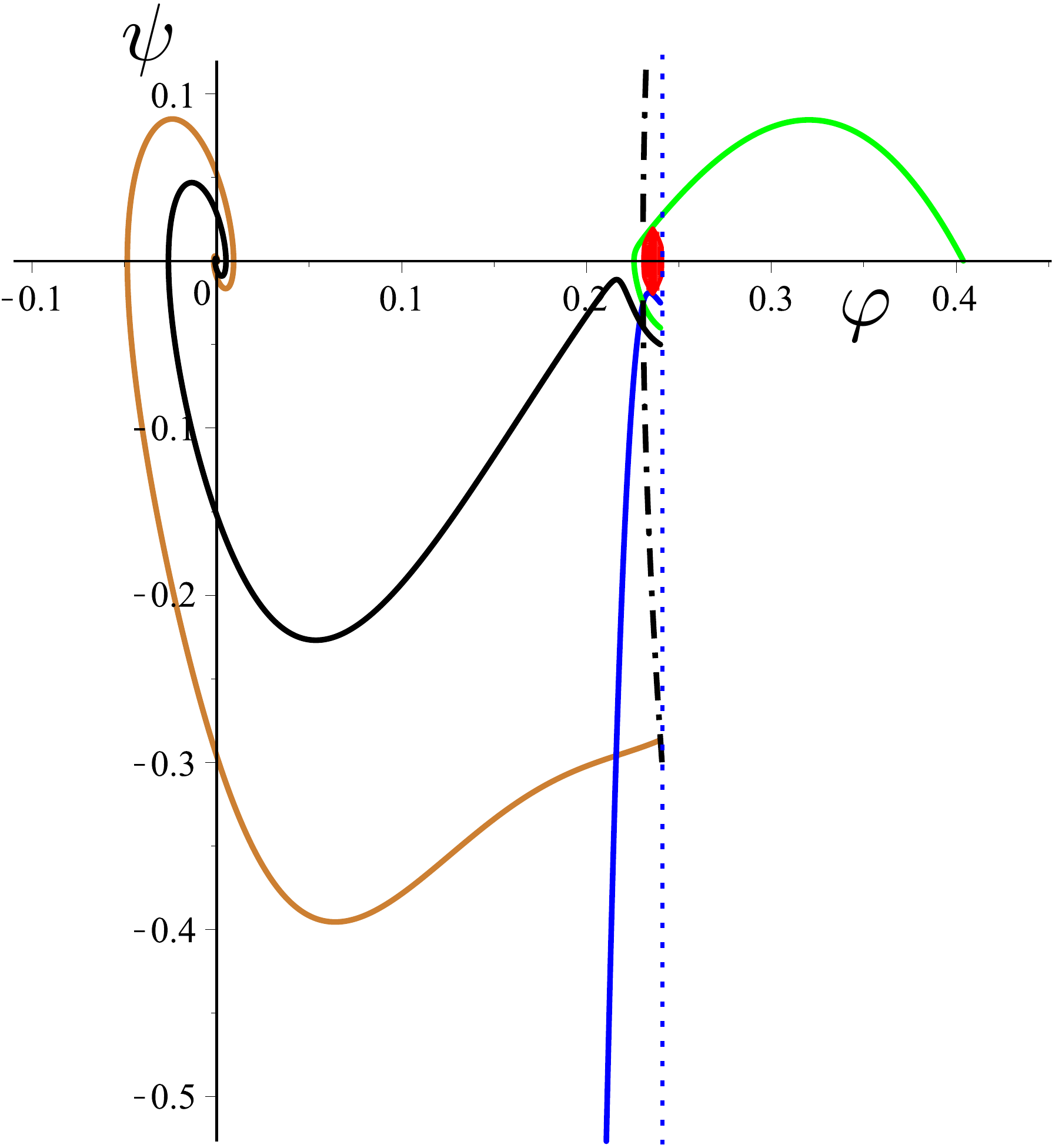} \qquad
\includegraphics[width=45mm]{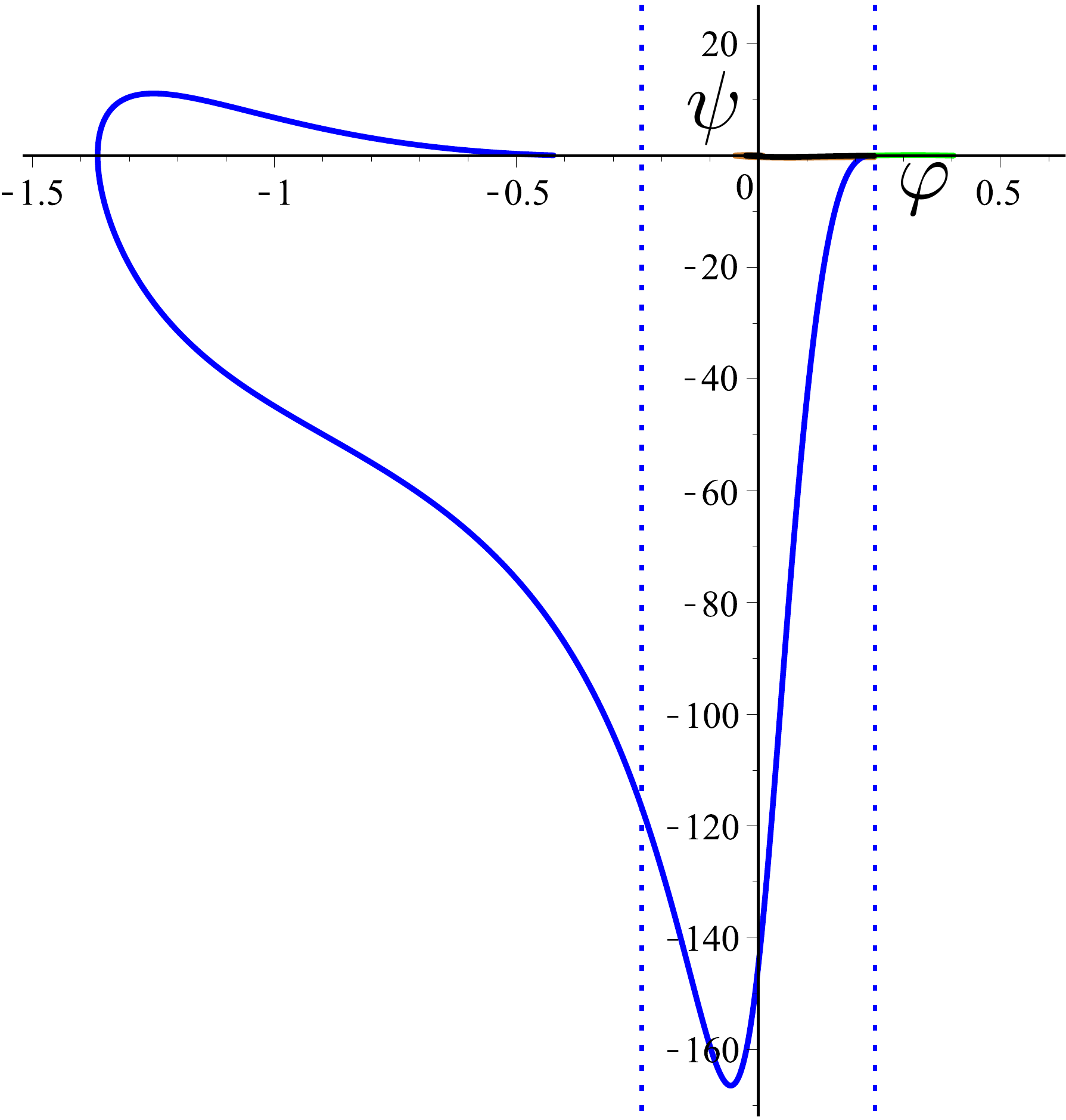}
\caption{There are basins of attraction for three possible attractors for the case of $V=C_6\varphi^6+C_0$, $U=U_0-\xi\varphi^2/2$. In the left picture: de Sitter solutions (black circles), trajectories with a bounce for $G_{eff}>0$ and then going to the antigravity region (green circles) and trajectories without a bounce for $G_{eff}>0$ going to the antigravity domain (blue circles). The parameters are $n=6$, $\xi=0.86$, $K=20$, $C_6=-1000$, and $C_0=0.15$. The phase trajectories (middle and right pictures) corresponding these attractors are plotted for initial data $\varphi_i=0.24$ and $\psi_i=-0.025$ (blue curve), $\varphi_i=-0.04$ (green curve), $\psi_i=-0.05$ (black curve). The gold trajectory starts at the bounce point $\psi_i=-0.2867158035$. The black dash-point line corresponds to $H=0$. The blue point lines correspond to $U=0$. The red color in the left and middle pictures indicates the unreachable domains. The critical value of $\varphi=\sqrt{\frac{1}{K\xi}}\approx0.2411$.}
\label{V6ksi086}
\end{figure}

We explore numerically cases with $n=4$ and $n=6$, applying the first order system of differential equations \eqref{FOSEQU}. Initial data are chosen $0<\varphi_b\leqslant\varphi(0)<\sqrt{\frac{1}{K\xi}}$, ~~$\psi_b\leqslant\psi(0)<0$,   and real $H(0)<0$ is calculated as $H_+(0)$ (see \eqref{hpm}). Values of the scalar field $\varphi_b$ and its time derivative $\psi_b$ are at the bounce line ${\psi_b}^2=-2V(\varphi_b)$ (remind that we denote $\psi=\dot\varphi$). In Figs.~\ref{n4xi01767},~\ref{n4xi01}, and \ref{V6ksi086} parameters $K=20$ and $C_0=0.15$ are chosen for all these plots and, therefore, stable de Sitter solution is $\varphi_{dS}=0$, $H_{dS}=\sqrt{\frac{C_0 K}{3}}=1$.

\begin{figure}[!h]
\centering
\includegraphics[width=45mm]{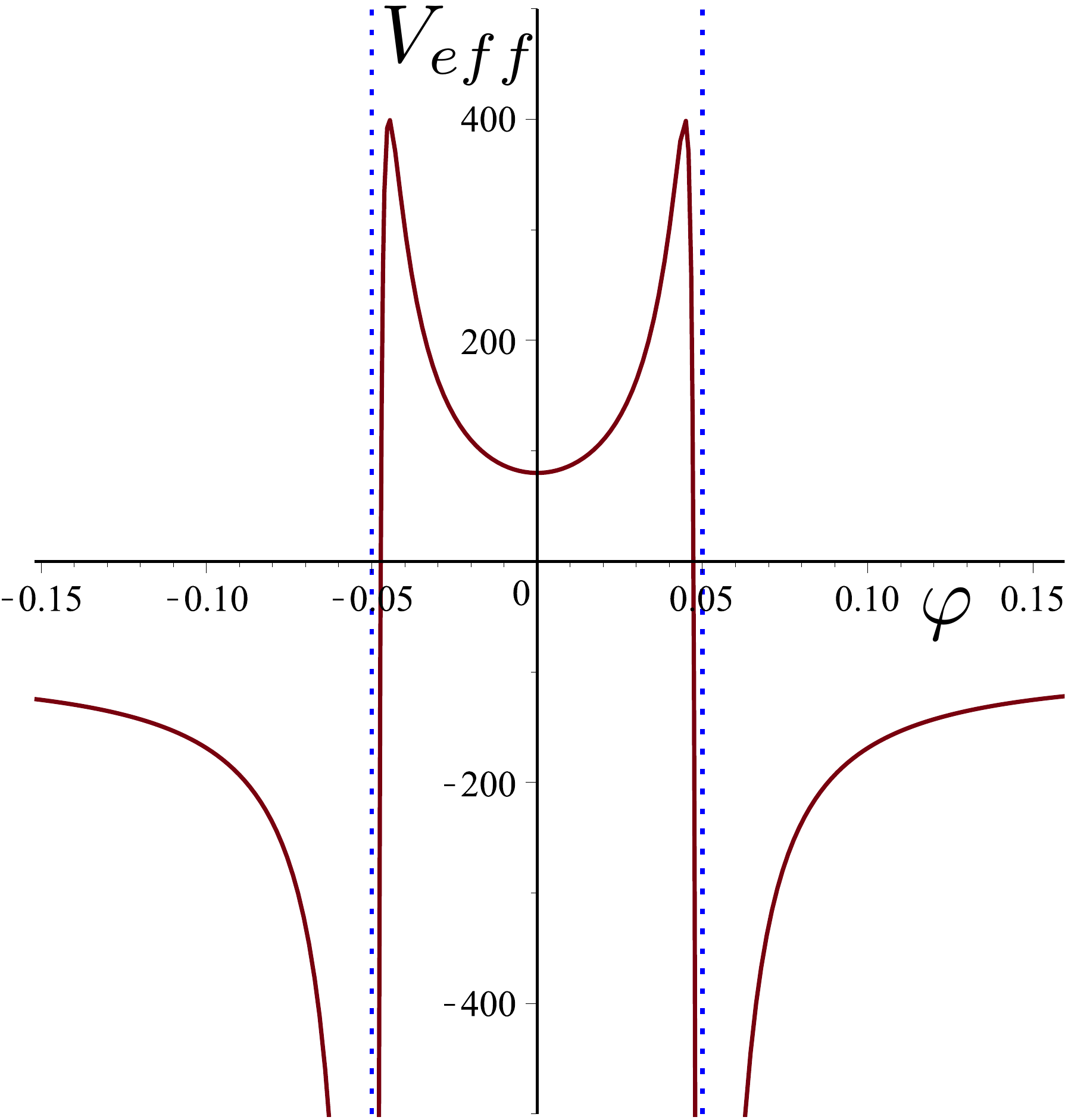}\qquad
\includegraphics[width=45mm]{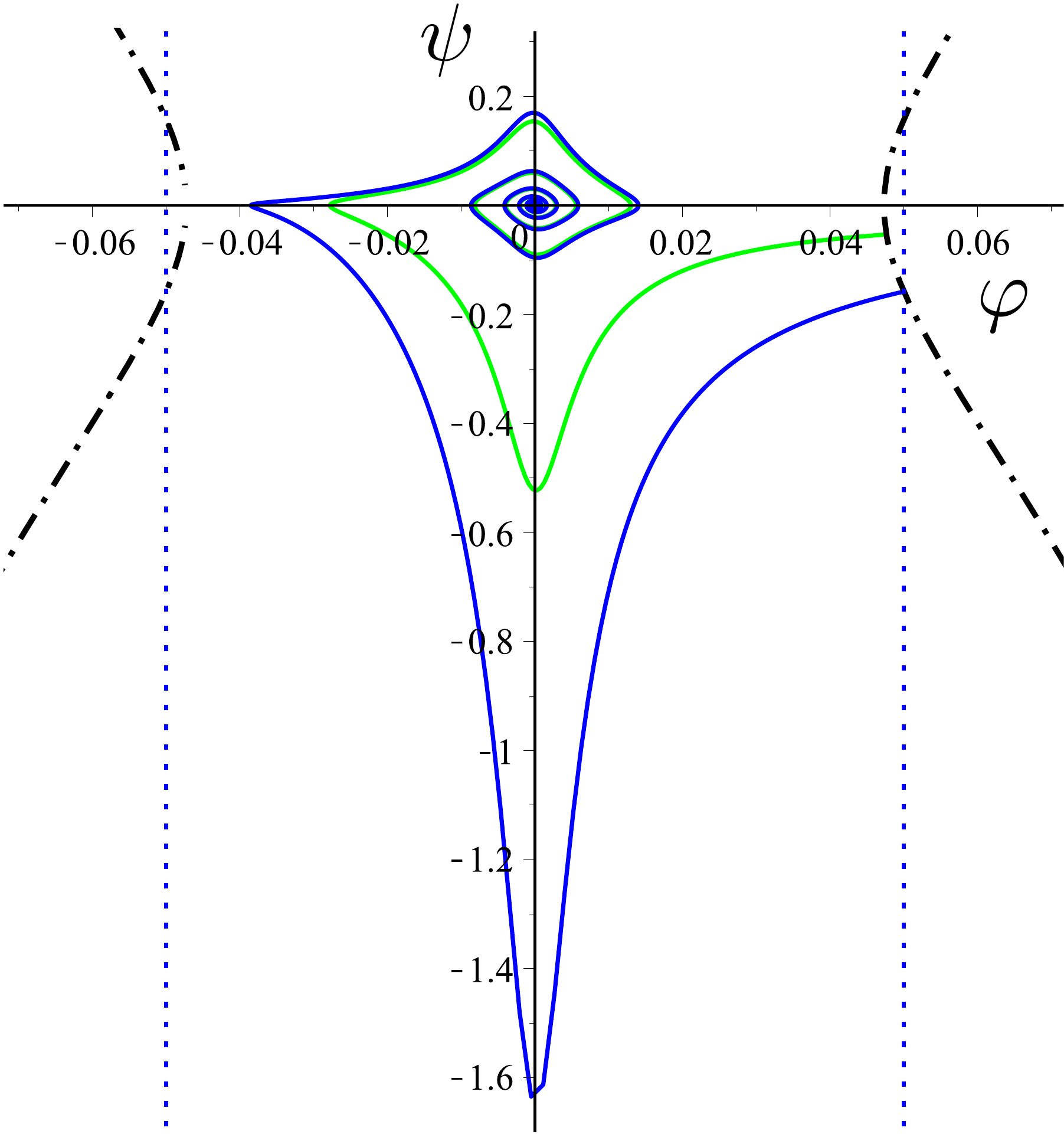}\qquad
\includegraphics[width=45mm]{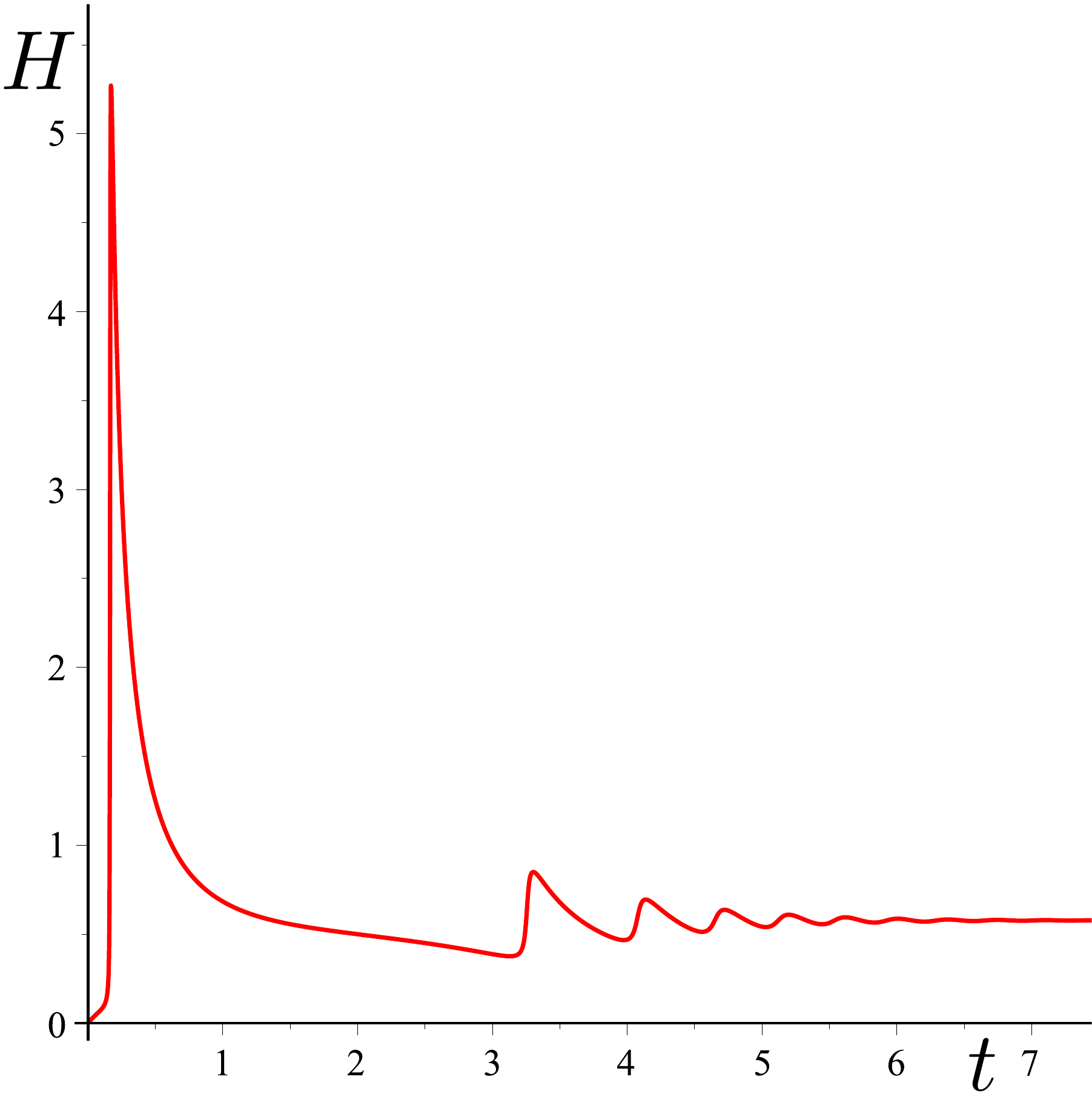}
\caption{The effective potential (left picture), phase trajectories (middle picture) and the Hubble parameter $H(t)$ (right picture) for $V=C_4\varphi^4+C_0$, $U=U_0-\xi\varphi^2/2$. The parameters are $\xi=20$, $K=20$, $C_4=-10000$, $C_0=0.05$. The initial conditions are $\varphi_i=1/21$ and $\psi_i=-0.05327109254$ (green line), $\varphi_i=0.04999750012$ and $\psi_i=-0.1580348161$ (blue line).  The black curves are the lines of the points that correspond to $H=0$. The blue point lines correspond to $U=0$.}
\label{n4xi20}
\end{figure}

The case of $n=4$ and $\xi=1/6$ is an integrable model with monotonic behavior of the Hubble parameter only. The change of parameter $\xi$ can not only give a Hubble parameter with one maximum (see~~\cite{Boisseau:2016pfh}), but also oscillating Hubble parameter that tends to a constant. Such a solution has been found at $\xi=20$ and is presented in Fig.~\ref{n4xi20}.

What is more interesting, numerical integration reveal another possible fate of a trajectory with initial conditions in the region suitable for bounce. In Fig.~\ref{n4xi01767} the outcome of the evolution with given initial values has been marked by different colors. Black and green zones correspond to a bounce followed by a smooth evolution towards de Sitter attractor (black) or $\varphi$-turn then growing $\varphi$, leading to crossing $U=0$ boundary (green). However, the blue zone exists starting from which a trajectory does not experience any bounce and fall into a singularity. This happens because a trajectory can touch the unreachable part of phase space and jump to another branch of two possible solutions for $H$ (see (\ref{hpm})). We have described such a phenomenon in the previous section, where this happens near $\varphi=0$ due to a negative cosmological constant $C_0$. Here it happens in the bounce zone, where we also have negative potential and unreachable part of the phase space. After branch changing, the trajectory cannot go through a bounce because bounce region for other branch is located in different phase space region. One can see in Fig.~\ref{n4xi01} that de Sitter attractors do not exist for sufficiently small positive $\xi$.


 Finally, let us mention the case of sixth degree potentials. We consider $\xi<\xi_{cr}\approx0.9410$. In Fig.~\ref{V6ksi086} one can see three possible behavior of the potential. The basins of attraction and phase trajectories demonstrate that solutions that do not leave the domain $U>0$ are similar to the case of the fourth order potential.
 One can see it in Fig.~\ref{n4xi01}, comparing the presented basins of attractions in the cases $n=4$ and $n=6$.
 There is the only difference between solutions that cross the line $U=0$ (see green and blue curves in Fig.~\ref{V6ksi086}), but we do not  consider features of behaviour of solutions in the antigravity domain in this paper.

The basins for the case of $n=2$ are qualitatively the same as for the $n=4$ and $n=6$ cases, so
we do not present them here. It is interesting that despite we got an additional analytical
restriction for the parameter $\xi$ to get de Sitter solution after the bounce, specific
only for $n=2$, it does not appear in numerics. The reason is that when decreasing $\xi$ the
de Sitter solution after the bounce disappears earlier, the way is similar to $n=4$ and $n=6$ cases.

\section{Stability of solutions}
In previous sections we have analyzed the stability of solutions in the FLRW metric. Note that the considering models include a scalar field with a standard kinetic term, so the bouncing solutions that do not leave the domain $U>0$ are free of the ghost instability~\cite{Gannouji:2006jm}.

The stability of solutions of the most general scalar--tensor theories has been investigated in~\cite{DeFelice:2011bh}. We use the results of this paper to analyze the stability of the solutions obtained. In the notations of~\cite{DeFelice:2011bh} action~(\ref{S1}) is
\begin{equation}
\label{S1h}
    S=\int d^4x\sqrt{-g}\left(G_4(\varphi)R+K(\varphi,X)\right),
\end{equation}
where
\begin{equation*}
G_4=U,\quad K=X-V(\varphi),\quad X=- \frac12\partial_\alpha\varphi\partial^\alpha\varphi=\frac12\dot{\varphi}^2.
\end{equation*}
Also,
\begin{equation*}
w_1=2U,\qquad w_2=4UH+2U'\dot\varphi,
\qquad
w_3=\frac{3}{2}\dot{\varphi}^2-18H^2U-18HU'\dot\varphi,\qquad w_4=2U.
\end{equation*}

The no-ghost condition for scalar perturbations
\begin{equation}\label{Qs}
Q_S\equiv\frac{w_1(4w_1w_3+9w_2^2)}{3w_2^2}>0
\end{equation}
 leads to
\begin{equation}
\begin{array}{l}
\label{noghostcond}
w_1(4w_1w_3+9w_2^2)=2 U \left( 8 U\left( \frac{3}{2}{\dot\varphi}^2-18 H^2 U-18 H \dot U\right) +9
{(4 H U+2 \dot U)}^2\right) =\\
=24 U{\dot\varphi}^2\left(U+3(U')^2\right)>0.
\end{array}
\end{equation}
This condition is satisfied for all trajectories that are in the gravity domain ($U>0$).
Note that $w_2\neq 0$ is equivalent to  $U a^2\neq \mathrm{const}$.

Let us consider the speed of propagation
\begin{equation}
\label{cS2}
c_S^2\equiv\frac{3(2w_1^2w_2H-w_2^2w_4+4w_1\dot{w}_1w_2-2w_1^2\dot{w}_2)}{w_1(4w_1w_3+9w_2^2)}.
\end{equation}

The denominator has been calculated in (\ref{noghostcond}) and is positive at $U>0$.
Using $w_4=w_1>0$, we rewrite the condition $c_S^2\geqslant 0$ as follows:
\begin{equation}
2w_1w_2H-w_2^2+4\dot{w}_1w_2-2w_1^2\dot{w}_2\geqslant 0.
\end{equation}
Taking into account Eq.~(\ref{Frequ21}),  we calculate
\begin{equation}
\label{noinstabcond}
2w_1w_2H-w_2^2+4\dot{w}_1w_2-2w_1^2\dot{w}_2= 4 \dot{\varphi}^2\left(U+3(U')^2\right)\geqslant 0.
\end{equation}

After substitution of the expressions (\ref{noghostcond}) and (\ref{noinstabcond}) in (\ref{cS2}) we get that
\begin{equation}
c_S^2=1.
\end{equation}

Conditions for the avoidance of ghosts and Laplacian instabilities for tensor perturbations are
\begin{equation*}
\begin{split}
Q_T&\equiv\frac{w_1}{4}=\frac{U}{2}>0 \quad \Rightarrow \quad U>0,\\
 c_T^2&\equiv\frac{w_4}{w_1}=1> 0.
\end{split}
\end{equation*}

Thus, all conditions that are necessary for the consistency of the model are satisfied for any trajectory that lie in the gravity domain ($U>0$). The obtained bouncing solutions do not suffer from the Laplacian (gradient) or ghost instability.

\section{Conclusion}

In this paper we have considered a class of spatially flat FLRW models with bounce solutions, generalizing those found in~\cite{Boisseau:2015hqa}. We show that despite the model with the potential $V_{int}(\varphi)=C_0+C_4 \varphi^4$, studied in~\cite{Boisseau:2015hqa}, is the unique model with the curvature scalar $R$ being the integral of motion, a much wider class of models admits bounce solutions followed by smooth future asymptotic behavior.

One class of models, studied in the present paper, generalizes the above-mentioned potential $V_{int}(\varphi)$ by adding a quadratic term. This generalization allows us to find bounce solutions with non-monotonic behavior of the Hubble parameter. The above-mentioned integrable model has a stable de Sitter solution at $\varphi=0$ that is a stable node. We have shown that the case of a stable focus at $\varphi=0$ is also possible. In this case we found solutions with essential decreasing Hubble parameter, including the solutions with the Hubble parameter tending to zero.

Moreover, we present a list of possible types of the bounce solutions and show that bounce behavior is still present even in the case when $C_0$ being negative, though in this case the nice future asymptotic is lost. In the case $C _0< 0$ there is an  unreachable domain on the phase plane in the neighborhood of the point $(0,0)$. Any point inside this domain corresponds to a  non-real value of the Hubble parameter. By this reason a trajectory can not crosses the boundary of this domain. Numerical calculations show that trajectories revolve around this domain. At the first glance some phase trajectories can revolve around this unreachable domain infinitely long, but this  is not the case. Analyzing the behavior of the function $P$, we prove that any trajectory that revolves around the unreachable domain touches it an a finite moment of time. After this moment the trajectory starts to tends to antigravity domain with   $U<0$.

Another generalization of the integrable  model replaces the quartic term in the potential by other even index. We studied the quadratic, quartic and sixth-power cases, changing also the coupling constant $\xi$. We show that despite the model becomes non-integrable, the qualitative features of solution found in~\cite{Boisseau:2015hqa} are still valid for not so restrictive conditions on the potential. At the same time, some new interesting features like possible branch changing during the cosmological evolution appears, which can decrease the measure of initial conditions good for bounce, though does not eliminate the possibility for the bounce completely. Only for sufficiently small positive $\xi$ the bounce solution that tends to de Sitter point does not exist.

We have analyzed the stability of the solutions obtained. All obtained bouncing solutions that tend to a stable de Sitter solution at $\varphi=0$ are free of the gradient or ghost instability.

   Inflationary models with the Ricci scalar multiplied by a function of the scalar field are very popular~\cite{nonmin-infl,HiggsInflation,Kaiser,Kallosh:2013hoa,Bamba:2014daa,EOPV2014,Pieroni2015}. They not only do not contradict the recent observational data~\cite{Planck2015}, but also connect cosmology and particle physics. Models with non-minimally coupled scalar fields are quite natural because quantum corrections to the effective action with minimal coupling include induced gravity terms~\cite{ChernikovTagirov,Callan:1970ze}.

  Note that the Einstein frame model that corresponds to the bouncing model~\cite{Boisseau:2015hqa} describes inflation~\cite{Bars1} (see also~\cite{Roest}).
    Also, non-local model~\cite{Biswas:2005qr} and its generalization~\cite{Koshelev:2013lfm} not only has bounce solution, but also (at other values of parameters) can describe inflation~\cite{Koshelev:2016xqb}.

    The observation data~\cite{Planck2015} are consistent with ekpyrotic cyclic models in which the universe is smoothed and flattened during a period of slow contraction followed by a bounce (see~\cite{Ijjas:2015hcc} and references therein). For certain matter bounce scenarios  in which the universe starts with a
matter-dominated contraction phase and transitions into an ekpyrotic phase the values of the spectral index and of the running parameter are in agreement with the observations~\cite{Cai:2014bea,Elizalde:2014uba}. At the same time there exists  a "no-go" theorem that a single field matter bounce model starting with vacuum initial conditions for the fluctuations  is ruled out by
observations~\cite{Quintin:2015rta}. To get scale-invariant perturbations with a slight red tilt
and a small tensor-to-scalar ratio models with contracting universe composed of cold dark matter, radiation, and a positive cosmological constant (the so-called $\Lambda$CDM bounce) have been proposed~\cite{Cai:2014jla} (see~\cite{Cai:2016hea} as a review). Note also that there exists the $f(R)$ gravity description of a $\Lambda$CDM bouncing model, without the need for matter fluids or for cosmological constant~\cite{Odintsov:2015zua}.

It would be very interesting to construct cosmological model with a non-minimal coupling standard scalar field, a bounce solution of which is suitable for inflationary scenario. The present paper can be considered as a step in this direction. We plan to consider a possibility to construct such a solution in future publications. In distinguish to the model proposed in~\cite{Wan:2015hya} we do not plan to include an additional Galileon term in our model.

Research of E.P. is supported in part by grant MK-7835.2016.2  of the President of Russian Federation.
A.T. is supported by RFBR grant 14-02-00894 and by the Russian Government
Program of Competitive Growth of Kazan Federal University.
Research of S.V. is supported in part by grant NSh-7989.2016.2 of the President of Russian Federation.
Researches of E.P. and S.V. are supported in part by the RFBR grant 14-01-00707.

\end{document}